 \documentclass[11pt]{article}

 \pdfoutput=1

\usepackage{graphicx} % for .epsi formats
\usepackage{amsmath}

\usepackage{enumitem}

\usepackage{hyperref}

\usepackage[francais,english]{babel}

\usepackage{float}
\usepackage{ifthen}
\begin{document}

%-------------------------------------------------------------------

\title{On a general definition of the squared Brunt-V\"{a}is\"{a}l\"{a} Frequency associated with the specific moist entropy potential temperature.}

\author{by Pascal Marquet and Jean-Fran\c{c}ois Geleyn. {\it M\'et\'eo-France}}

%\corraddr{Pascal MARQUET, DPr\'evi/Labo, M\'et\'eo-France, 42 av. G. Coriolis, 31057 Toulouse CEDEX 01, France.\\ Web site: http://perso.numericable.fr/$\sim$pmarquet/  ; E-mail: pascal.marquet@meteo.fr }

%\date{\today}
\date{14th of January, 2014}

\maketitle

%-------------------------------------------------------------------

\begin{center}
{\em Paper submitted in July 2011 to the \underline{Quarterly Journal of the Royal Meteorological Society}. }\\
{\em (Published: Volume 139, Issue 670, pages 85-100, January 2013 Part A).} \\
{\em http://onlinelibrary.wiley.com/doi/10.1002/qj.1957/abstract.} \\
{\em Additional changes are included in Appenxix~F. They correspond to a note published} \\
{\em in the \underline{WGNE Blue-book} in 2012: ``Moist-entropy vertical adiabatic lapse rates:} \\
{\em the standard cases and some lead  towards inhomogeneous conditions''.} \\
{\em \underline{Corresponding addresses}: pascal.marquet@meteo.fr / jean-francois.geleyn@meteo.fr}
\end{center}
\vspace{1mm}

%-------------------------------------------------------------------

\begin{abstract}
The squared Brunt-V\"{a}is\"{a}l\"{a} Frequency (BVF) is computed in terms of the moist entropy potential temperature recently defined in Marquet (2011).
Both homogeneously saturated and non-saturated versions of $N^2$ (the squared BVF) are derived.
The method employed for computing these special homogeneous cases relies on the expression of density written as a function of  pressure, total water content and specific moist entropy only.
The associated conservative variable diagrams are discussed and compared with existing ones.
Despite being obtained without any simplification, the formulations for $N^2$ remain nicely compact and are clearly linked with the squared BVF  expressed in terms of the adiabatic non-saturated and saturated lapse rates.
As in previous similar expressions, the extreme homogeneous solutions for $N^2$ are of course different, but they are not analytically discontinuous.
This allows us to define a simple bridging expression for a single general shape of $N^2$, depending only on the basic mean atmospheric quantities and on a transition parameter, to be defined (or parameterized) in connection with the  type of application sought. 
This integrated result remains a linear combination (with complex but purely local weights) of two terms only, namely the environmental gradient of the moist entropy potential temperature and the environmental gradient of the total water content. Simplified versions of the various equations are also proposed for the case in which the moist entropy potential temperature is approximated by a function of both so-called moist-conservative variables of Betts (1973).
\end{abstract}

%-------------------------------------------------------------------

% -----------------------
\section{Introduction.} % (Section 1)
% -----------------------
\label{section_INTRO}

Several attempts have been published to express the  Brunt-V\"{a}is\"{a}l\"{a} Frequency (hereafter BVF) in terms of the two conservative variables represented by the total water specific content (i.e. for closed systems) and the specific moist entropy (i.e. for closed, reversible and adiabatic systems).
The methods described in Durran and Klemp (1982) and  Emanuel (1994) -- hereafter referred to as DK82 and E94, respectively --  mainly differ by the choice of the ``moist entropy'' formulation to be used as a moist conservative variable.

The entropy potential temperature $\theta_s$ recently defined in Marquet (2011, hereafter referred to as M11) corresponds to a general formulation for the specific moist entropy, valid for any parcel of moist atmosphere with varying specific content of dry air, water vapour and liquid or solid water.

The  aim of this article is therefore to derive non-saturated and saturated versions of the squared BVF expressed in terms of $\theta_s$, and to compare them comprehensively with the previous moist formulations published in DK82 and E94.
In this respect, the objective of the article is more general than that targeted in Geleyn and Marquet (2010), where the objective was to express the squared BVF in terms of an approximate formulation for $\theta_s$, namely $(\theta_s)_1$.
Comparisons between the exact and the approximate versions of the specific moist entropy defined in terms of $\theta_s$ and $(\theta_s)_1$ will be realized with the help of the conservative variable diagrams published in Pauluis (2008, 2011, hereafter referred to as P08 and P11).

The paper is organized as follows.
The mathematical definition of the moist squared BVF ($N^2_m$) is presented in section~2, with $N^2_m$ expressed  in terms of the gradients of the two conservative variables ($s, q_t$) and established in the Appendix~B.
The moist definition of the state equation is recalled in the section~3, together with M11's  specific moist entropy defined in terms of $\theta_s$.
The non-saturated and saturated versions $N^2_{ns}$ and  $N^2_{sw}$ are then presented in sections~4 and 5, with some detailed  computations available in the Appendices~C and D, respectively.

Several comparisons between $N^2_{ns}$ and $N^2_{sw}$ and the previous versions published in DK82 and E94 are presented in the section~6.
The comparisons are made either with the formulation expressed in terms of the lapse rate formulation or in terms of the gradients of the two conservative variables ($s, q_t$), with a special attention paid to the latter one in the section~7.
The non-saturated and saturated approximate versions of the moist squared BVF expressed with $(\theta_s)_1$ are presented in the section~8.

Some numerical applications are presented in the section~9 with the use of the same FIRE-I data sets than in M11. 
Separate analyses are made for in-cloud (saturated) and clear-air (non-saturated) air.
Additional analyses are presented in the Appendix~E and F.
First, the non-saturated moist squared BVF is compared with the usual formula expressed in terms of the vertical gradient of the virtual potential temperature.
Second, the possibility of defining an analytic transition between the non-saturated and the saturated versions of the moist squared BVF is explored, using a control parameter $C$ varying continuously between $0$ and $1$.
Finally, conclusions are presented in section 10.

% ---------------------------------------------
\section{The moist squared Brunt-V\"{a}is\"{a}l\"{a} Frequency} % (Section 2)
% ---------------------------------------------
\label{section_BVF}

It is shown in Appendix~B that the moist squared BVF can be defined by
\begin{align}
  N^2_{m}  & = \: - \: \frac{g}{\rho} \:
                \left( \;
                  \left.
                  \frac{\partial \rho}{\partial s}
                  \right|_{p,q_t} \:
                \frac{\partial  s  }{\partial z  }
                \:\; + \:\;
                  \left.
                  \frac{\partial \rho}{\partial q_t}
                  \right|_{p,s} \:
                \frac{\partial q_t }{\partial z  }
                \; \right)
  \: . \label{def_new_N2_moist_6}
\end{align}
Formulation (\ref{def_new_N2_moist_6}) is different from the classical one used  in DK82 or E94, for instance.
It is assumed that the density can be expressed as a function of the two conserved variables $s$ and $q_t$, as well as of pressure $p$, leading to $\rho = \rho(s, q_t , p)$.
The moist squared BVF can then be expressed as a weighting sum of the two local vertical gradients of $s$ and $q_t$, with weighting factors depending on appropriate partial derivatives of the density with respect to $s$ and $q_t$.

In order to compute the moist formulation (\ref{def_new_N2_moist_6}), the density must be expressed analytically in terms of the three independent variables $(s,q_t,p)$.
One of the problems is that such an explicit formulation for $\rho(s,q_t,p)$ does not exist for the moist air.
It is however possible to define $\rho$ in terms of $(s,q_t,p)$ by expressing the entropy and the state equations as  $\rho(T,q_t,p)$ and $s(T,q_t,p)$, and then by eliminating the temperature.
This method is used in the sections \ref{sec_N2m_unsat} and \ref{sec_N2m_sat} and in Appendices C and D to compute the non-saturated and saturated version of the squared BVF.

The saturated case is more difficult  to deal with than the non-saturated one, because of the existing condensed water terms $q_l$ or $q_i$.
But these terms can be expressed as differences between $q_t$ and the saturated values $q_{sw}$ or $q_{si}$, which depend on $T$,  $p$ and $q_t$.
It is thus necessary to express the temperature $T$ in terms of $(s,q_t,p)$, at least implicitly.

% ---------------------------------------------
\section{The state and the entropy equations} % (Section 3)
% ---------------------------------------------
\label{section_entropy_state_Eqs}

The state equation of the moist air is written with the use of the dry gas constant $R_d$ replacing the moist value $R = q_d\:R_d + q_v\:R_v$, yielding
\begin{align}
  p  &  =
\: \rho \ R \: T \:
\; = \:
\: \rho \ R_d \: T_v \:
  \: , \label{def_eq_state}
\end{align}
where the virtual temperature $T_v$ (Lilly, 1968) is equal to
\begin{align}
  T_v  & \; = \:
  T \: \left(\: 1 \: + \: \delta \:q_v \: - \: q_l \: - \: q_i \: \right)
  \: = \: \left( \frac{T}{\theta} \right)  \;  {\theta}_{v}
  \: , \label{def_Tv2} \\
  T_v  &  =
\: T \:
  \left( \: 
  \frac{1 + \eta \: r_v}
        {1+r_t}
  \: \right)
  \; = \:
  T \: \left(\: 1 \: + \: \eta \: q_v \: - \: q_t\: \right)
  \: . \label{def_Tv1}
\end{align}
From the Appendix~A, the constants are equal to $\delta = R_v/R_d-1\approx 0.608$, and $\eta = \delta+1$.
The virtual temperature corresponds to  Eq.(9) in DK82 and to the ``density temperature'' denoted by  $T_{\rho}$ in E94.
The associated (liquid water) virtual potential temperature is equal to $ {\theta}_{v}$ defined in (\ref{def_Tv2}).

The specific moist entropy is defined in M11 as
\begin{equation}
  s  \; \equiv \: s_{ref}
                + {c}_{pd} \:
  \ln\left( {\theta}_{s} \right)
    , \,
  \label{def_S_THs}
\end{equation}
where the reference entropy $s_{ref}$ is equal to
\begin{equation}
  s_{ref}  \; = \: s_d^0 \:
                - {c}_{pd} \:
        \ln\left( T_0 \right)
  \; \approx 1138.56 \mbox{~J~K}^{-1} \mbox{~kg}^{-1}
  \label{def_Sref}
\end{equation}
and where the entropy potential temperature ${\theta}_{s}$ can be written as
\begin{align}
  {\theta}_{s} 
  & =
        \: \theta \;
        \exp \left( - \:
                    \frac{L_{vap}\:q_l + L_{sub}\:q_i }{{c}_{pd}\:T}
                \right)
      \; \exp \left( \Lambda_r \:q_t \right) \;
  \nonumber \\
  & \quad \times \;
        \left( \frac{T}{T_r}\right)^{\lambda \:q_t}
        \left( \frac{p}{p_r}\right)^{-\kappa \:\delta \:q_t}
    \left(
      \frac{r_r}{r_v}
    \right)^{\gamma\:q_t}
     \;
      \frac{(1+\eta\:r_v)^{\:\kappa \: (1+\:\delta \:q_t)}}
          {(1+\eta\:r_r)^{\:\kappa \:\delta \:q_t}}
  \: . \label{def_THs}
\end{align}
The advantage of the definition (\ref{def_S_THs}) for $s$, with $c_{pd}$  and $s_{ref}$ being  two numerical constants,  is that ${\theta}_{s}$ becomes trully synonymous with or the specific moist entropy, whatever the thermodynamic properties of the parcel (i.e. $T$, $p$, ..., $q_t$, $q_l$, $r_v$, ...) may be.

It can be verified (see M11) that both $s_{ref}$ and ${\theta}_{s}$ are independent of the reference values $T_r$ and $p_r$, providing that $r_r(T_r,p_r)$ is equal to the saturating values $\varepsilon\:/\:[\;p_r/e_{sw}(T_r)-1\:]$ or $\varepsilon\:/\:[\;p_r/e_{si}(T_r)-1\:]$, depending on $T_r>T_0$ or $T_r<T_0$, respectively.
All the different terms of $s_{ref}$ and ${\theta}_{s}$ depend on $T_r$ and $p_r$ in such a way that $s_{ref}$ and ${\theta}_{s}$ remain unchanged.

The new term $\Lambda_r$ appearing in (\ref{def_THs}) is the main difference from previous studies on moist entropy and associated moist potential temperatures.
It depends on the difference between the reference partial entropy values of dry air and water vapour,  leading to the numerical value $\Lambda_r = [ (s_{v})_r - (s_{d})_r ] / c_{pd} \:  \approx \: 5.87$ (as computed in M11 from the Third law and for the reference state, see Appendix~A for the reference values of entropies).

The impact of the reference values of partial entropies on the definition of the moist air entropy has been addressed independently in Pauluis et al. (2010, hereafter referred to as PCK10), where an equivalent of $\Lambda_r$ has been studied in the Appendices, in the form of an unknown and arbitrary constant ``$a$''.

In constrast, the moist entropy is computed in E94 and P11 ``per unit mass of dry air'', with the reference values $(s_{d})_r$ and $(s_{v})_r$ suppressed from the formulae of the specific values $s_d$ and $s_v$.
This method prevents the possibility of arriving at a relevant definition for the specific moist entropy (``per unit mass of moist air''), with varying values for $q_d$ or $q_t=1-q_d$ to be put in factor of $(s_{d})_r$ and $(s_{v})_r$, respectively.

The three terms in the right-hand side of the first line of (\ref{def_THs}) correspond to the variables $\theta_l$ and $q_t$ defined in Betts (1973, hereafter referred to as B73), with $q_t$ multiplied by $\Lambda_r$.
The third term in the second line of (\ref{def_THs}) might be irrelevant for the dry-air limit where $q_t=q_v$ and $r_v=q_v/(1-q_v)$ tends to $0$ when $q_v$ tends to $0$.
However, this term varies like $q_v \,\ln(q_v)$ which has the limit $0$ when $q_v$  tends to $0$.

%==================================
\section{The squared BVF for the unsaturated moist air.}    %  (Section  4)
%==================================
\label{sec_N2m_unsat}

The unsaturated moist air is defined by $q_l=q_i=0$ and $q_t=q_v$.
From (\ref{def_S_THs}) and since $c_{pd}$ is a constant, the gradient of the specific moist entropy is exactly equal to
\begin{align}
  \frac{\partial s}{\partial z}
  & = \:
      c_{pd}\:
      \frac{\partial \ln(\theta_s) }{\partial z} 
  \: . \label{def_dsdz}
\end{align}

The computations of the two partial derivatives of the density involved in the formulation  (\ref{def_new_N2_moist_6}) for $N_m^2$ are described in the Appendix~C.
They are given by (\ref{def_ns_drho_ds_2}) and (\ref{def_ns_drho_dqt_3}), leading to
\begin{align}
 &  N^2_{ns}
   = \;
    \Gamma_{ns} \;
      \frac{\partial s }{\partial z}
              \:
\: + \:  
       g \:
     \frac{\partial \ln(q_d)  }{\partial z}
\: + \: 
      \Gamma_{ns}
  \left[ \:
    (1+r_v) \: \frac{c_p \: R_v}{R}
    \: - \:
   c_{pd} \:
 \left(
            \Lambda_r + \Lambda_v
            \right)
    \: \right]
      \frac{\partial q_v  }{\partial z}
              \:
  \: .
\label{def_BVF_ns_new}
\end{align}

The term $\Lambda_v$ and the unsaturated adiabatic lapse rate $\Gamma_{ns}$ are equal to
\begin{align}
  \Lambda_v
  & \: = \:
  \lambda \: \ln\left( \frac{T}{T_r} \right)
  \: - \:
  \kappa \: \delta \: \ln\left( \frac{p}{p_r} \right)
  \: - \:
  \gamma \: \ln\left( \frac{r_v}{r_r} \right)
  \: + \:
  \kappa \: \delta \:
      \ln\left( \frac{1+\eta\:r_v}
                      {1+\eta\:r_r}
      \right)
  \: , \label{def_ns_Lambda_v} \\
  \Gamma_{ns}
  &    = \:
     \frac{g}{c_p}
        \: = \:  - \: 
                  \left.
                  \frac{\partial T}{\partial z}
                  \right|_{s,q_v}
    = \:  
                 \frac{p \: g}{R\:T}
                  \left.
                  \frac{\partial T}{\partial p}
                  \right|_{s,q_v}
  \: .
\label{def_BVF_G_ns}
\end{align}
The derivative of $T$ with respect to pressure is computed in (\ref{def_BVF_G_ns}) at constant specific moist entropy and water content.
The resulting value $\Gamma_{ns}$, with $c_p$ attaining its moist value depending on $q_v$, has been obtained after long computations involving all the terms entering in the specific moist entropy formulation defined by (\ref{def_S_THs})-(\ref{def_THs}), in a way analogous to the computations described in some details in the Appendix~C.

The  term in (\ref{def_BVF_ns_new}) involving the gradient of $q_d$ can be written as any of the alternative ways
\begin{align}
  - \:  
       \frac{ g }{1-q_v} \:
     \frac{\partial q_v  }{\partial z}
 &  
  = \: 
     g  \:
     \frac{\partial \ln(1-q_v)  }{\partial z}
  = \:  
     g  \:
     \frac{\partial \ln(q_d)  }{\partial z}
\label{def_BVF_ns_new_last1} \\
 &
  = \: - \:  
       \frac{ g }{1+r_v} \:
     \frac{\partial r_v  }{\partial z}
  \: .
\label{def_BVF_ns_new_last2}
\end{align}
The formulation involving $q_d$ corresponds to the original one derived in Lalas and Einaudi (1974, hereafter referred to as LE74) for the saturated case.
It is retained for the present non-saturated version in order to be consistent with the next section.

It is important to notice that only the hydrostatic approximation has been made to derive (\ref{def_BVF_ns_new}), according to the demonstration given in the Appendix~B to obtain the squared BVF formulation (\ref{def_new_N2_moist_6}), where the specific moist entropy is defined by the more general and exact formula (\ref{def_S_THs}), with ${\theta}_{s} $ given by almost all the terms in (\ref{def_THs}), except that $q_l=q_i=0$ and $q_t=q_v$.

%==================================
\section{The squared BVF for the saturated moist air.}    %  (Section  5)
%==================================
\label{sec_N2m_sat}

The computations of the two partial derivatives of the density involved in the saturated formulation  (\ref{def_new_N2_moist_6}) for $N_m^2$ are described in the Appendix~D.
They are given by (\ref{def_sat_drho_ds_2}) and (\ref{def_sat_drho_dqt_2}) and the liquid-water saturated counterpart  of (\ref{def_BVF_ns_new}) writes
\begin{align}
  & N^2_{sw}
   = \;  \Gamma_{sw} \;
      \frac{\partial s }{\partial z}
\: + \: g \:
      \frac{\partial \ln(q_d)  }{\partial z}
\: + \: \Gamma_{sw}
  \left[ \:
     (1+r_{sw}) \: \frac{L_{vap}}{T}
      -  c_{pd}
            \left(
            \Lambda_r + \Lambda_{sw}
            \right)
     \: \right]
      \frac{\partial q_t  }{\partial z}
  \: .
\label{def_BVF_sat_new_l}
\end{align}
The terms $\Lambda_{sw}$ and $\Gamma_{sw}$ are equal to
\begin{align}
  \Lambda_{sw}
  & \: = \:
  \lambda \: \ln\left( \frac{T}{T_r} \right)
  \: - \:
  \kappa \: \delta \: \ln\left( \frac{p}{p_r} \right)
  \: - \:
  \gamma \: \ln\left( \frac{r_{sw}}{r_r} \right)
  \: + \:
  \kappa \: \delta \:
      \ln\left( \frac{1+\eta\:r_{sw}}
                          {1+\eta\:r_r}
      \right)
  \: , \label{def_sat_Lambda_s}
\\  \Gamma_{sw}
  &   = \:
     \frac{g}{c_p}
     \;
     \frac{D_{1w}}{D_{2w}}
     \: = \: 
                 \frac{p \: g}{R\:T}
                  \left.
                  \frac{\partial T}{\partial p}
                  \right|_{s,q_t}
 \: ,  \label{def_BVF_G_sw}
\end{align}
where
\begin{align}
  D_{1w}
  & = \: 1 \: + \:
      \left( 1 + \eta \: r_{sw} \right) \:
      \frac{L_{vap} \: q_{sw}}{R_{d}\:T_v}
  \: , \label{def_sat_D1l} \\
  D_{2w}
  & = \: 1
  \:
  + \:
      \left( 1 + \eta \: r_{sw} \right) \:
      \frac{L_{vap}^2 \: q_{sw}}{c_p\:R_{v}\:T^2}
  \: . \label{def_sat_D2l}
\end{align}

The liquid-water saturated adiabatic lapse rate is given by (\ref{def_BVF_G_sw}), with the derivative of $T$ with respect to pressure computed at constant specific moist entropy and water content.
The resulting value $\Gamma_{sw}$ has been obtained after long computations involving the specific moist entropy defined by (\ref{def_S_THs}) and  (\ref{def_THs}), in a way analogous to the one described in some details in  Appendix~D.
The specific heat $c_p$ is the moist version of it, depending on $q_d$, $q_v$ and $q_l$, as expressed by (\ref{def_prop_sat}) or in  Appendix~A.

The term in (\ref{def_BVF_sat_new_l}) involving the gradient of $q_d$ can be written in any of the alternative ways
\begin{align}
  - \:  
       \frac{ g }{1-q_t} \:
     \frac{\partial q_t  }{\partial z}
 &  
  = \: 
     g  \:
     \frac{\partial \ln(1-q_t)  }{\partial z}
  = \:  
     g  \:
     \frac{\partial \ln(q_d)  }{\partial z}
\label{def_BVF_sat_new_last1} \\
 &
  = \: - \:  
       \frac{ g }{1+r_t} \:
     \frac{\partial r_t  }{\partial z}
  \: .
\label{def_BVF_sat_new_last2}
\end{align}
The last formulation (\ref{def_BVF_sat_new_last2}) is used in DK82 and E94.
The one involving $q_d$ is the original one derived in Eq.(43) of LE74, where the superscript ``$1$'' represents the dry air density.
The corresponding term was written in the following way, due to the property $q_d = \rho_d / \rho$.
\begin{align}
  g \: \frac{\partial \ln(q_d)}{\partial z}
 &  = \: - \: g \: 
 \left[ \:
  \frac{\partial \ln(\rho)}{\partial z}
  \: - \: 
  \frac{\partial \ln(\rho_d)}{\partial z}
 \: \right]
  \: .
\label{def_BVF_sat_new_last3}
\end{align}

As for the unsaturated case, only the hydrostatic  approximation has been made to derive (\ref{def_BVF_sat_new_l}).
In particular, the specific moist entropy is defined by the more general and exact formula (\ref{def_S_THs}) and ${\theta}_{s} $  by  (\ref{def_THs}), with most of the terms varying with $s$, $q_t$ or $p$.

The ice-water saturated counterparts  of (\ref{def_BVF_sat_new_l}) to (\ref{def_sat_D2l})  are obtained by replacing $L_{vap}$ by $L_{sub}$,  $r_{sw}$ by $r_{si}$ and $q_{sw}$ by $q_{si}$, including in the moist definition of $c_p$.

In comparison with the non-saturated formulation (\ref{def_BVF_ns_new}) and the associated weighting factor (\ref{def_ns_Lambda_v}), the saturated formulation (\ref{def_BVF_sat_new_l}) for $N^2$ is modified so that the term $\Lambda_v$ must be replaced by its saturated equivalent $\Lambda_{sw}$ given by (\ref{def_sat_Lambda_s}), with  ${c_p}\: {R_v}/{R}$ replaced by $L_{vap}/T$ and with the  term (${D_{1w}}/{D_{2w}}$)  appearing in the adiabatic lapse rate (\ref{def_BVF_G_sw}) equal to $1$ in the unsaturated adiabatic lapse rate formulation (\ref{def_BVF_G_ns}).

%==================================
\section{Full comparisons with DK82 and E94.}    %  (Section  6)
%==================================
\label{sec_N2m_other_BVF2}

In order to better compare the present results with those published in DK82 and E94, the unsaturated and saturated squared BVF  formulations given by (\ref{def_BVF_ns_new}) and (\ref{def_BVF_sat_new_l}) can be rewritten in terms of the gradient of temperature, by computing the vertical derivative of all the terms entering the formulation for  $\ln(\theta_s)$, with $\theta_s$ given by  (\ref{def_THs}).
It appears that most of the terms cancel out, leading to
\begin{align}
 N^2_{ns}
   & = \;
       \frac{ g }{T} \:
  \left(
      \frac{\partial T }{\partial z}
      +
      \Gamma_{ns} 
  \right)
  \: + \:  
     g \: \delta \:
     \frac{ T }{T_v} \:
     \frac{\partial q_v  }{\partial z}
  \: ,
\label{def_BVF_ns_for_DK82} \\
 N^2_{sw}
   & = \;
       \frac{ g \:D_{1w}}{T} \:
  \left(
      \frac{\partial T }{\partial z}
      +
      \Gamma_{sw} 
  \right)
  \: - \: 
     \frac{ g }{1+r_t} \:
     \frac{\partial r_t  }{\partial z}
  \: .
\label{def_BVF_sat_for_DK82}
\end{align}

The saturated squared BVF given by Eq.(13) in DK82 can be rewritten, with the notation of the Appendix~A, as
\begin{align}
 N^2_{DK}
   & = \;
       \frac{ g \:D_{1w}}{T} \:
  \left(
      \frac{\partial T }{\partial z}
      +
      \Gamma_{DK} 
  \right)
  \: - \: 
     \frac{ g }{1+r_t} \:
     \frac{\partial r_t  }{\partial z}
  \: ,
\label{def_BVF_sat_DK82}
\end{align}
with the same formulation for $D_{1w}$ as in (\ref{def_sat_D1l}), due to the equality
\begin{align}
      1 \: + \: 
      \left( 1 + \eta \: r_{sw} \right) \:
      \frac{L_{vap} \: q_{sw}}{R_{d}\:T_v}
 & = \:
      1 \: + \: 
      \frac{L_{vap} \: r_{sw}}{R_{d}\:T}
  \: . \label{def_sat_D1_DK82}
\end{align}
It is worthwhile to notice that all the mixing ratios were denoted by the letter ``$q$'' in DK82, and that the use of the more standard letter ``$r$'' is made in the present article.

The difference between (\ref{def_BVF_sat_for_DK82}) and (\ref{def_BVF_sat_DK82}) concerns the  liquid-water saturated adiabatic lapse rate (\ref{def_BVF_G_sw}), defined by Eq.(19) in DK82, leading to
\begin{align}
 \Gamma_{\!DK}
  &   = \:
     \frac{g}{c_{pd}}
     \; (1+r_t) \: 
     \frac{D_{1w}}{D_{DK}}
 \: .  \label{def_G_DK}
\end{align}
The lapse rate computed in DK82 contains an additional term $(1+r_t)$ and $g/c_p$ is replaced by $g/c_{pd}$.
Moreover, the term at the denominator of (\ref{def_G_DK}) may be written as
\begin{align}
  D_{DK} & = \: 1
  \: + \:
      \left( 1 + \eta \: r_{sw} \right) \:
      \frac{L_{vap}^2 \: r_{sw}}{c_{pd}\:R_{v}\:T^2}
  \: + \:
      \frac{ c_{pv}\:  r_{sw} + c_{l} \: r_{l}}{c_{pd}}
  \:  . \label{def_D_DK}
\end{align}
It is different from $D_{2w}$ given by (\ref{def_sat_D2l}) in that $c_p$ is replaced by $c_{pd}$, $q_{sw}$ by $r_{sw}$, with the additional last term depending on $r_{sw}$ and $r_{l}$.
All these differences make the formulations  (\ref{def_BVF_sat_for_DK82}) and (\ref{def_BVF_sat_DK82}) more unlike than what could appear at a first sight.

The absence of the last term of (\ref{def_D_DK}) in the $\theta_s$ formulation for $D_{2w}$ given by (\ref{def_sat_D2l}), and the direct multiplication of the unsaturated adiabatic gradient by $D_{1w}/D_{2w}$  in  (\ref{def_BVF_G_sw}), indicate that the use of the $\theta_s$ formulation eventually leads to a more compact and more logical definition of the saturated adiabatic lapse rates.

The  liquid-water saturated adiabatic lapse rate is defined in E94 by
\begin{align}
 \Gamma_{\!EM}
  &   = \:
     \frac{g}{c^{\ast}_{p}}
     \; (1+r_t) \: 
     \frac{D_{1w}}{D_{EM}}
 \: .  \label{def_G_EM94}
\end{align}
The E94 formulation contains the same additional term $(1+r_t)$ than in the DK82 formulation (\ref{def_G_DK}),  but with $c_p$ replaced by $c^{\ast}_{p} =  c_{pd}  + c_{pv} \:  r_{sw}$ instead of $ c_{pd}$.
The term at the denominator may be written as
\begin{align}
  D_{EM} & = \: 1
  \:
  + \:
      \left( 1 + \eta \: r_{sw} \right) \:
      \frac{L_{vap}^2 \: r_{sw}}{c^{\ast}_{p}\:R_{v}\:T^2} 
  \: + \:
      \frac{ c_{l} \: r_{l}}{c^{\ast}_{p}}
  \:  . \label{def_D_EM94}
\end{align}
In comparison with the $\theta_s $  formulation $D_{2w}$ given by (\ref{def_sat_D2l}), the E94's formulation $D_{EM}$ contains the additional third term, with $q_{sw}$ replaced by $r_{sw}$ in the second term, and $c_p$ replaced by $c^{\ast}_{p}$ in the second and third terms.

The equivalent of (\ref{def_BVF_sat_for_DK82}) or (\ref{def_BVF_sat_DK82}) is not mentioned in E94.
However Emanuel, like Durran and Klemp, tried to express the moist and saturated value of the squared BVF in terms of conservative variables, with the same quantity $q_t$ representing the conservation of the dry air or the total water species, but with definitions of the moist entropy that are different from the one retained in the present paper, depending on $\theta_s$ given by  (\ref{def_THs}).

The moist entropy-like function appearing in DK82 was expressed in terms of the quantity $c_{pd} \: \ln(\theta_q)$, with  $\theta_q$ defined by
\begin{align}
 \theta_q
  & = \:
  \theta_E
  \left(
    \frac{T}{T_0} 
 \right)^{c_l\:r_t/c_{pd}}
\: , \label{def_Theta_q} \\
 \theta_E 
  & = \: 
  \theta \:
  \exp \left(
    \frac{L_{vap}\:r_{sw}}{c_{pd}\:T} 
 \right)
\; . \label{def_Theta_E}
\end{align}
It is worthwhile noting that, from (\ref{def_Theta_q}) and provided the correction term depending on $(T/T_0)$ is a small one (valid in the lower troposphere where $T \approx T_0$ and above where $r_t$ tends to $0$), $\theta_q$ is almost equivalent to the equivalent potential temperature $\theta_E$ given by (\ref{def_Theta_E}).

The corresponding saturated value of the squared BVF is given by Eq.(21) in DK82.
It writes
\begin{align}
 N^2_{DK}
   & = \;
     \frac{\Gamma_{\!DK} }{1+r_t} \:
       \: c_{pd} \:
      \frac{\partial \ln(\theta_q) }{\partial z}
      - \:  \frac{ g }{1+r_t} \:
     \frac{\partial r_t  }{\partial z}
  \: .
\label{def_BVF_sat_DK82_bis}
\end{align}
The lapse rate $\Gamma_{\!DK}$ is given by (\ref{def_G_DK}) and (\ref{def_D_DK}).

The moist entropy-like function appearing in E94 will be denoted by $s^{\ast}$ in the present paper.
It is different from the $\theta_s$ specific entropy formulations (\ref{def_S_unsat}) and (\ref{def_S_sat})  in that Emanuel, like P11, has considered an entropy ``per unit mass of dry air'' and not ``per unit mass of moist air''.
As a consequence, the reference values are not derived from the Third Law in E94 and P11, and the corresponding term $\Lambda_r$ does not appear as such.

More precisely, the non-saturated and saturated versions of the moist entropy $s^{\ast} = s/q_d = s/(1-q_t) = (1+r_t)\:s$ are defined in E94 by
\begin{align}
  s^{\ast}
   & = \:
     (c_{pd}  + c_{pv} \:  r_t) \: \ln(T)
    \: - \: R_d \: \ln(p_d)
    \: + \: \frac{L_{vap} \: r_v}{T}
    \: - \: r_v \: R_v\: \ln(e/e_{sw})
  \: .
\label{def_BVF_S_star_EM94}
\end{align}
The liquid-water saturated version of (\ref{def_BVF_S_star_EM94}) is obtained by replacing $ r_v$ by $ r_{sw}$ and with a relative humidity of $100$\% leading to $e/e_{sw}=1$, and therefore to a cancellation of the last term.
It is possible to compare the saturated version $s^{\ast}_{sw}$  with (\ref{def_S_THs})-(\ref{def_THs}) by using the properties $p_d = p/(1+\eta\:r_v)$ and $r_{sw}=r_t-r_l$, leading to
\begin{align}
s^{\ast}_{sw} & \:\equiv\: \:
     c_{pd} \:\ln(\theta^{\ast} / p_0^\kappa)
     \label{def_BVF_sat_S_star_EM94_bis} \\
\theta^{\ast}
      & \:=\:
        \: \theta \;
      \; \exp \left( - \frac{L_{vap} }{{c}_{pd}\:T} \: r_l \right)
      \; \exp \left( \frac{L_{vap} }{{c}_{pd}\:T} \: r_t \right) \;
      \;\;  T^{(1+\lambda) \:r_t} \;\; {(1+\eta\:r_{sw})^{\:\kappa }}
  \: . \label{def_sat_THs_star_EM94}
\end{align}
The general features of $\theta_s$ and $\theta^{\ast}$ are similar.
The  main difference between (\ref{def_THs}) and (\ref{def_sat_THs_star_EM94}) is that the term $\Lambda_r \approx 5.87$ in $\theta_s$ is replaced by ${L_{vap} }/{({c}_{pd}\:T)}\approx 9$ in the second exponential of (\ref{def_sat_THs_star_EM94}).
Moreover, the specific contents are replaced by the mixing ratios and the second and third lines of (\ref{def_THs}) are different from the last two terms in (\ref{def_sat_THs_star_EM94}).

The  saturated squared BVF corresponding to $s^{\ast}_{sw}$ is derived in E94.
It is equal to
\begin{align}
 N^2_{EM}
   & = \:
      \frac{\Gamma_{\!EM} }{1+r_t} \:
      \frac{\partial s^{\ast} }{\partial z} 
      \: - \: 
     \frac{ g }{1+r_t} 
     \frac{\partial r_t  }{\partial z}
      \: - \:
      \frac{\Gamma_{\!EM}}{1+r_t} \:  
  \left[ \: c_l \: \ln(T) \: \right] \:
     \frac{\partial r_t  }{\partial z}
\label{def_BVF_sat_EM94_bis} \: ,
\end{align}
with $\Gamma_{\!EM}$ given by (\ref{def_G_EM94}) and (\ref{def_D_EM94}).

The comparisons  between  (\ref{def_BVF_sat_DK82_bis}) or (\ref{def_BVF_sat_EM94_bis}) and the present formulation (\ref{def_BVF_sat_new_l}) show the following:
\begin{itemize}[label=-,leftmargin=3mm,parsep=0cm,itemsep=0cm,topsep=0cm,rightmargin=2mm]
\item  the last terms $-(1+r_t)^{-1}{\partial r_t  }/{\partial z}$ in the DK82 and E94 formulations and the term ${\partial \ln(q_d)  }/{\partial z}$ in (\ref{def_BVF_sat_new_l}) are the same, due to the properties (\ref{def_BVF_sat_new_last1})-(\ref{def_BVF_sat_new_last2});
\item  the moist lapse rates $\Gamma_{\!DK}$ and $\Gamma_{\!EM}$ are different from the one (\ref{def_BVF_G_sw}) obtained with $\theta_s$, as explained above;
\item  both $\Gamma_{\!DK}$ and $\Gamma_{\!EM}$ are divided by $(1+r_t)$ in (\ref{def_BVF_sat_DK82_bis}) and (\ref{def_BVF_sat_EM94_bis}), removing the impact on $N^2_{\!DK}$ and $N^2_{\!EM}$ of the same extra term $(1+r_t)$ included in these adiabatic lapse rates; 
\item  the moist entropy functions are not the same, with $s$ and $c_{pd} \:  \ln(\theta_s)$ different from  $s^{\ast}$ or $c_{pd} \:  \ln(\theta_q)$, implying vertical gradients different of those in the E94 and DK82 formulations (\ref{def_BVF_sat_DK82_bis}) and (\ref{def_BVF_sat_EM94_bis});
\item  the last bracketered term of (\ref{def_BVF_sat_new_l}), which represents a new term consistent with the new formulation for the specific moist entropy and $\theta_s$, does not appear in the DK82 formula (\ref{def_BVF_sat_DK82_bis}), and is only partially present in the E94 formula (\ref{def_BVF_sat_EM94_bis}).
\end{itemize}

The formulation (\ref{def_BVF_sat_EM94_bis}) has been  expressed in E94 with the hope of managing vertical gradients of conservative variables only.
It has been concluded that ``cloudy air is stable if moist entropy increases upward and total water decreases upward''.
This is  true only if the moist entropy is accurately represented by $s^{\ast}$ in E94.
The same property holds for the DK82 formulation (\ref{def_BVF_sat_DK82_bis}), as far as the moist entropy is accurately represented by  $c_{pd} \: \ln(\theta_q)$.

A similar stability analysis may be applied to the present formulation (\ref{def_BVF_sat_new_l}), with the same stable feature valid  for $N^2_{sw}$ on the first line of (\ref{def_BVF_sat_new_l}) if $\theta_s$  and $q_d$ increase upward,  but with the presence of a non-negligible contribution in the second line, of the opposite sign to ${\partial \ln(q_d)  }/{\partial z} > 0$ and larger in absolute value for classical atmospheric conditions.
The novel aspects of this contribution, without any equivalent in the DK82's and E94's formulations (\ref{def_BVF_sat_DK82_bis}) and (\ref{def_BVF_sat_EM94_bis}), will be studied in detail in the next section.

The second line of (\ref{def_BVF_sat_new_l}) almost disappears in E94 and does not exist  in DK82.
It is possible to explain this feature by comparing the potential temperature $\theta_s$ given by  (\ref{def_THs}) with the E94 formulation $\theta^{\ast}$ given by (\ref{def_sat_THs_star_EM94}).
Clearly, the term $\Lambda_r$ in  $\theta_s$  is replaced by ${L_{vap} }/{({c}_{pd}\:T)} \approx 9$ in $\theta^{\ast}$.
This modification would transform the second line in (\ref{def_BVF_sat_new_l}) into a term depending on ${r}_{sw} \: L_{vap} \, / \, T - {c}_{pd}\: {\Lambda}_{sw} $.
It is a residual quantity which is much smaller than $L_{vap} \, / \, T - {c}_{pd}\: {\Lambda}_{r}$ and this explains why the impact of the new term ${\Lambda}_{r}$ depending on the absolute values of the reference partial entropies is  important in the second line of  (\ref{def_BVF_sat_new_l}) for  $N^2_{sw}$.

%==============================================================================
\section{Impact of gradients of $q_t$ and of moist entropy formulations.}    %  (Section  7)
%==============================================================================
\label{sec_N2m_grad_qt}

The  remarks and comparisons mentioned in the previous section indicate that the DK82 and E94 formulations possess two potential conceptual drawbacks with respect to the present proposal.

First, the DK82 and E94 moist lapse rates (\ref{def_G_DK}) and (\ref{def_G_EM94}) should not contain the term $(1+r_t)$.
This term is a consequence of the moist entropy being defined ``per unit mass of dry air'' in DK82 and E94, corresponding to the transformation of any specific value $\psi$ into $\psi^{\ast} = (1+r_t)\:\psi= \psi/(1-q_t)= \psi/q_d$.
It is important to notice  that, even if the moist lapse rates are multiplied by $(1+r_t)$, this has no impact on $N^2_{DK}$ and $N^2_{EM}$, since the lapse rates are divided by the same factor  in (\ref{def_BVF_sat_DK82_bis}) and (\ref{def_BVF_sat_EM94_bis}).
The adiabatic lapse rates of a parcel in a moist atmosphere should however get a unique definition, and the specific moist entropy based on $\theta_s$ and (\ref{def_THs}) appears to be a more general way to obtain it, leading to the simple shape of $\Gamma_{sw}$ given by (\ref{def_BVF_G_sw})-(\ref{def_sat_D2l}).

Second, it seems that one of the constraints for determining or choosing the moist entropy formulations $s^{\ast}/q_d$ or ${c}_{pd}\:\ln(\theta_q)$ may be to express the squared BVF as a sum of two terms only, with a first term depending only on the gradient of moist entropy formulations plus another term depending only on the gradient of $q_t$, but being exactly equal to the prescribed value $- \: g/(1+r_t) \: ({\partial r_t}/{\partial z})$ given by  (\ref{def_BVF_sat_new_last2}).
It is indeed the result (\ref{def_BVF_sat_DK82_bis}) obtained in DK82 and almost achieved in E94, where a partial second line still exists in (\ref{def_BVF_sat_EM94_bis}).

%===============
% Figure (1) :  % Pauluis_S_q_t /  T_rho_S_qt_v0.proc  / T_part1 
%===============
\begin{figure}[hbt]
\centering
 \includegraphics[width=0.55\linewidth,angle=0,clip=true]{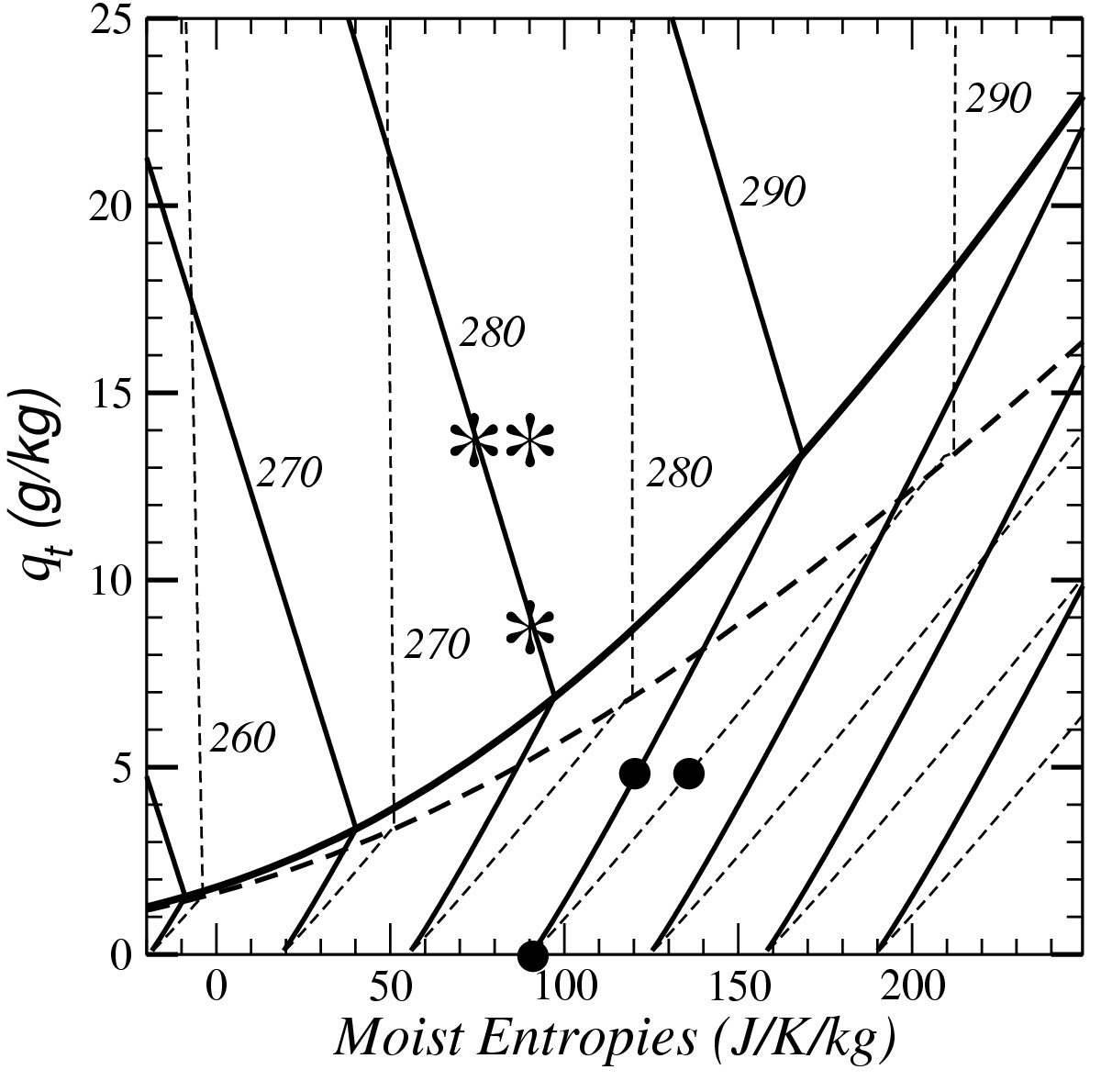}
\caption{\it\small
The same conservative variable diagram as in PS10 with total specific water content plotted against the moist entropies.
The temperature $T(s, q_t, p_1)$ is plotted for $p_1=900$~hPa.
The diagram is separated by the increasing saturation curves located in the center of it, with non-saturated and saturated regions located below and above the curves, respectively.
The dashed lines represent equal values of $T(s^{\ast}, q_t, p_1)$ computed with the PS10 moist entropy $s^{\ast}(\theta^{\ast})$ given by  (\ref{def_BVF_S_star_EM94})-(\ref{def_sat_THs_star_EM94}), depending on the non-saturated or saturated conditions, respectively.
The solid lines correspond to equal values of $T(s, q_t, p_1)$ computed with the specific moist entropy formulations $s(\theta_s)$ given by (\ref{def_S_unsat}) or (\ref{def_S_sat}), depending on the non-saturated or saturated conditions, respectively.
The $\theta_s$ entropy formulations is plotted as $s-s^0_{d}$  on the x-axis, with $s^0_{d}=6775$~J~K${}^{-1}$~kg${}^{-1}$.
The lines of equal temperature are labelled every $10$~K.
The black circles represent the changes in $s$ or $s^{\ast}$ due to the moistening of an initially  dry-air parcel of atmosphere, with the same increase of $q_t$ of $5$~g~kg${}^{-1}$ and undergoing isothermal processes following lines of equal value of either $T(s, q_t, p_1)$ or $T(s^{\ast}, q_t, p_1)$.
The black stars represent the changes in moist entropies due to the moistening of an initially saturated parcel by the same amount of $5$~g~kg${}^{-1}$ and undergoing the same kind of isothermal processes.
\label{fig_PS10_diagrams}}
\end{figure}

The specific moist entropy has a unique formulation since it is a thermodynamic state function.
In this respect, different moist entropy formulations must lead to different sets of curves in the conservative variable diagram plotted in Figure~\ref{fig_PS10_diagrams}.
The solid lines and the dashed lines coincide for the dry air limit at the bottom of the diagram, due to the global shift by the amount of the dry air standard value $s^0_{d}$ for $s$.

The differences between lines of equal value of $T$ become more and more important as $q_t$ increases.
The saturation curves are also different, depending of the use of $s(\theta_s)$ versus $s^{\ast}(\theta^{\ast})$ formulations.
The dashed lines are almost equal values of $s^{\ast}$ in the saturated region, whereas the specific moist entropy $s$ decreases as $q_t$ increases in the saturated region.
These differences demonstrate that the way in which the moist entropy is defined may generate important differences in the physical interpretation.

In fact, it is unlikely that some arbitrariness may exist in the possibility to change the formulation of the specific moist entropy.
It is the difference in the formulations of the moist entropy that generates the large impacts on the definition of the moist squared BVF and concerning the different ways to write the terms depending on the gradient of $q_t$.
Moreover, even if the squared BVF values eventually remain close to each other, different values for the changes in moist entropy correspond to different physical meanings.

In particular, for isothermal and isobaric transformations where $q_t$ increases by the same given amount of $5$~g~kg${}^{-1}$, the black circles and stars represent the changes in moist entropies associated with isothermal and isobaric transformations.
The changes in moist entropies are clearly different in PA10's version from the present one. 
In saturated conditions, the change of entropy almost cancels out with $s^{\ast}(\theta^{\ast})$, whereas the specific moist entropy exhibits a large decrease with $s(\theta_s)$.
It may be considered that the almost isentropic feature appearing for the isotherms above the saturation curve with the formulations of PA10 and E94 is a very special case that is not explained by the theory, nor supported by observations.

The reference values are determined in $s(\theta_s)$ from the Third Law and $\Lambda_r \approx 6$ in (\ref{def_THs}) is replaced by $L_{vap}/(c_{pd}\:T)\approx 9$ in (\ref{def_sat_THs_star_EM94}).
It is worthwhile noting that when $\Lambda_r$ is arbitrarily set to $9$ in the $\theta_s$ definition (\ref{def_THs}), the solid lines are almost superimposed  on the dashed lines in Figure~\ref{fig_PS10_diagrams} (result not shown).
This indicates that the third Law used to compute the special value of $\Lambda_r$ is a key part for the definition of $s(\theta_s)$, and that the logic of PA10 and E94 is of a different kind.

It is therefore important to justify the present formulation, with the above prescribed term separated from the other terms in the second line of (\ref{def_BVF_ns_new}) and (\ref{def_BVF_sat_new_l}) even if those also depend on the gradient of $q_t$.
In fact, this prescribed term was introduced first in LE74  as  a correction to older formulations,  in the form $g\: {\partial \ln(q_d)}/{\partial z}$  given by (\ref{def_BVF_sat_new_last3}).
This correction term also appears in the  DK82's moist version in the equivalent form (\ref{def_BVF_sat_new_last2}), where it was still considered as an additional term.

By analogy with the ideas published in Pauluis and Held (2002), this correction term may be interpreted as a modification to the vertical stability, represented here by $N^2_{sw}$, and corresponding to a conversion between the kinetic and the potential energy due to the work required to ensure the vertical transport of water species.

Hence, the correction term given by (\ref{def_BVF_sat_new_last2})  logically appears in the DK82 and E94 conservative variable formulations (\ref{def_BVF_sat_DK82_bis}) and (\ref{def_BVF_sat_EM94_bis}) and also in the first lines of the present formulations (\ref{def_BVF_ns_new}) and (\ref{def_BVF_sat_new_l}) for $N^2_{ns}$ and $N^2_{sw}$, respectively, but given by the equivalent form (\ref{def_BVF_sat_new_last3}).
The correction term must not be included in the specific moist entropy term, nor be regrouped  with the second line of (\ref{def_BVF_ns_new}) and (\ref{def_BVF_sat_new_l}).

The computation of the moist squared BVFs in terms of the vertical gradient of $\theta_v$ are derived in the Appendix~E, showing that the moist squared BVF based on the specific moist entropy $s(\theta_s)$ is not exactly proportional to ${\partial \theta_v }/{\partial z}$ for a moist but non-saturated atmosphere.

An interesting feature suggested by the second lines of (\ref{def_BVF_ns_new}) and (\ref{def_BVF_sat_new_l}), in which the moist squared BVF is based on the $\theta_s$ approach, is that a continuous transition may exist between the two unsaturated and saturated regimes.
An example of this kind of transition is described in the Appendix~F.

Summing up, the important new result derived in the present paper is that the specific moist entropy (and $\theta_s$) really verify the conservative property associated with the second principle.
The consequence is that second lines must exist in both the non-saturated version (\ref{def_BVF_ns_new}) and the saturated one (\ref{def_BVF_sat_new_l}).
The physical meaning for these extra terms can be found in the results obtained in M11, where the vertical profile for the specific moist entropy $s$ is observed to be almost a constant for marine stratocumulus, in spite of existing vertical trends for the B73 variables  $q_t$ and $\theta_l$.
These results are true for both clear-air (unsaturated) and in-cloud (saturated) moist regions.

This means that, at least for marine stratocumulus, the sign of $N^2_{ns}$ and $N^2_{sw}$ are not controlled by the vertical gradient of specific moist entropy, but almost entirely by the vertical gradient of $q_t$.
More precisely, the vertical profiles of $q_t$ impact on $N^2_{ns}$ and $N^2_{sw}$ not only via the ``water lifting'' contributions located in the first lines of (\ref{def_BVF_ns_new}) and (\ref{def_BVF_sat_new_l}), but also via the terms located in the second lines, where ``expansion work'' and ``latent heat'' effects are accompanied by the new impact corresponding to the pure entropy terms $\Lambda_r$, $\Lambda_{v}$ and $\Lambda_{sw}$.

%===============
% Figure (2) :  / FIRE_I / zprog_trace_9 ; THETA_allstars3 " ; LTYPG=24 / zprog_convert: ITYP_22f
%===============
\begin{figure}[hbt]
\centering
 \includegraphics[width=0.50\linewidth,angle=0,clip=true]{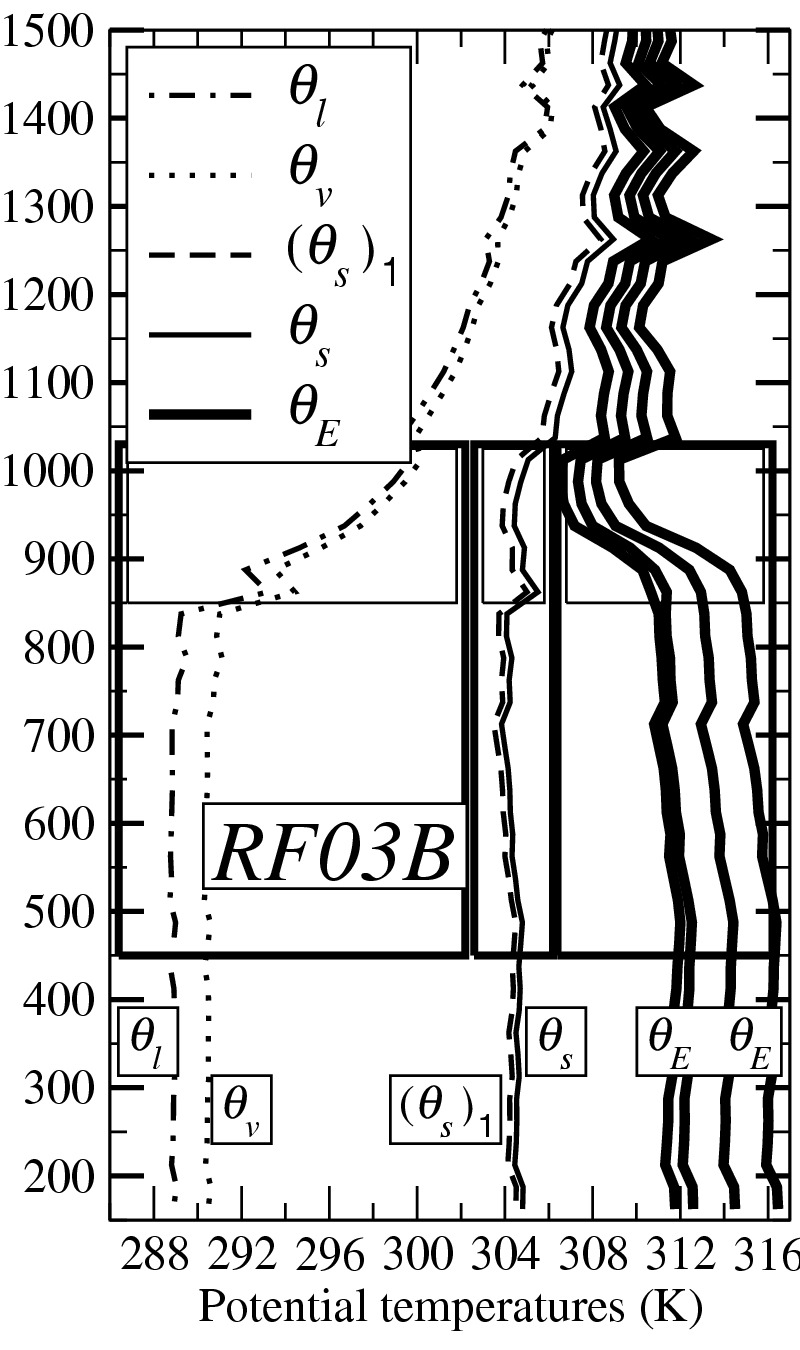}
\caption{\it\small
Comparison of ${\theta}_{s}$ (in a central position) with other potential temperature formulations for the grid-cell average of the RF03B profiles of the FIRE-I datasets.
The two profiles on the left are ${\theta}_{l}$ and  ${\theta}_{v}$ given by (\ref{def_THl}) and (\ref{def_Tv2}).
The four last profiles on the right correspond to four formulations of ${\theta}_{E}$ (B73, E94, Bolton (1980) and a formulation used in ARPEGE).
The vertical profile of the approximate version (\ref{def_THs1}) is located close to the  ${\theta}_{s}$ curve and is labelled by $({\theta}_{s})_1$.
\label{fig_THS1_part1}}
\end{figure}

The second lines of (\ref{def_BVF_ns_new}) and (\ref{def_BVF_sat_new_l}) are of opposite sign.
It is possible to explain this result by comparing 
\begin{equation}
\left[ \frac{R_v}{R} \approx 1.6 \right]
   \; < \: 
\left[ \; \Lambda_r  \approx 5.9 \; \right]
   \; < \: 
\left[ \frac{L_{vap}}{c_{pd}\: T} \approx 9  \right]
  \: .
\label{def_compare_1}
\end{equation}
The result is that $\Lambda_r$ is almost in a two-third / one-third position between ${R_v}/{R}$ and $L_{vap}/{(c_{pd}\: T) }$ in terms of the control parameter $C$ (see Appendix~F), leading to positive values for $L_{vap}/T - c_{pd}\: \Lambda_r$ in the second line of $N^2_{sw}$ and to negative values for the corresponding term in the second line of $N^2_{ns}$.
This result may be put into context with the property illustrated in  Figure~\ref{fig_THS1_part1}, where ${\theta}_{s}$ is almost in a two-third / one-third position between ${\theta}_{l}$ and  ${\theta}_{E}$ (see also M11).

%==================================
\section{Approximate versions for the moist squared BVF.}    %  (Section  8)
%==================================
\label{sec_N2m_Thetas1}

It is shown in M11 that the specific moist entropy defined by (\ref{def_S_THs}) can be accurately approximated by 
\begin{equation}
  s  \; \approx \: s_1 \; = \: s_{ref}
                + {c}_{pd} \:
  \ln\left[ ({\theta}_{s})_1 \right]
    . \,
  \label{def_S_THs1}
\end{equation}
The  reference entropy is still given by (\ref{def_Sref}), but the specific moist entropy potential temperature ${\theta}_{s}$  is approximated by $({\theta}_{s})_1$ given by the first line in the R.H.S. of  (\ref{def_THs}), leading to a liquid water version, which may be written as
\begin{align}
  ({\theta}_{s})_1
  & =
        \: \theta_l
       \; \exp \left( \Lambda_r \:q_t \right) 
  \: , \label{def_THs1}
\end{align}
where
\begin{align}
  \theta_l
  & =
        \: \theta \;
        \exp \left( - \:
                    \frac{L_{vap}\:q_l }{{c}_{pd}\:T}
                \right)
  \:  \label{def_THl}
\end{align}
is the B73 liquid-water potential temperature.

The specific moist entropy $s[(\theta_s)]$ and its approximate version $s_1[(\theta_s)_1]$ are compared in Figure~\ref{fig_THS1_part1} to other usual moist potential temperatures.
Clearly, the replacement of ${\theta}_{s}$ by $({\theta}_{s})_1$ is a good approximation, with errors $({\theta}_{s})_1 - {\theta}_{s}$ much smaller than the observed large differences between ${\theta}_{s}$ and ${\theta}_{l}$, ${\theta}_{v}$ or ${\theta}_{E}$.
Moreover, the errors are almost constant along the vertical and they should not largely impact on the computations of the moist squared BVF.

The  computations of the moist squared BVF and of the moist adiabatic lapse rates can be realized through the replacement of $s$ by $s_1$.
The corresponding approximate non-saturated and saturated versions of the squared BVF may be written as
\begin{align}
   N^2_{1/ns}
   & = \;
       \frac{ g }{c_{pd}} \:
      \frac{\partial s_1 }{\partial z}
              \:
\: + \:  g\:
     \frac{\partial \ln(q_d)  }{\partial z}
    \: + \: 
      g
  \left[ \:
    (1+r_v) \: \frac{R_v}{R}
    \: - \:
    \Lambda_r 
    \: \right]
      \frac{\partial q_v  }{\partial z}
              \:
  \: ,
\label{def_BVF1_ns_new}
\end{align}
and
\begin{align}
  N^2_{1/sw}
   & = \;  
     \frac{ g }{c_{pd}} \:
     \frac{ D_{1w} }{D_{2wl}} \:
      \frac{\partial s_1 }{\partial z}
\: + \:  g\:
     \frac{\partial \ln(q_d)  }{\partial z}
\: + \; g \:
     \frac{ D_{1w} }{D_{2wl}} \:
  \left[ \:
     (1+r_{sw}) \: \frac{L_{vap}}{c_{pd}\:T}
      -   \Lambda_r 
    \: \right]
      \frac{\partial q_t  }{\partial z}
  \: .
\label{def_BVF1_sat_new_l}
\end{align}
The term $D_{1w}$ is still given by (\ref{def_sat_D1l}) but $D_{2w}$ is replaced by
\begin{align}
  D_{2wl}
  & = \: 1
  \:  + \:
      \left( 1 + \eta \: r_{sw} \right) \:
      \frac{L_{vap}^2 \: q_{sw}}{c_{pd}\:R_{v}\:T^2}
  \: + \: 
  \frac{ L^0_{v}\:q_l}{c_{pd}\:T}
  \: . \label{def_sat_D2l1}
\end{align}
A new term depending on $q_l$ appears in $D_{2wl}$, with the specific heat  $c_p$ in (\ref{def_sat_D2l}) replaced by $c_{pd}$ in  (\ref{def_sat_D2l1}) and $L^0_{v}$ given by (\ref{def_Lv_0}).

The last term involving $L^0_{v}$ and $q_l$ is somehow similar to the last terms in the DK82 and E94 formulations (\ref{def_D_DK}) and (\ref{def_D_EM94}).
These terms are of the order of $1.8\:r_{sw}$ in $D_{DK}$, $4.2\:r_l$ in $D_{EM}$ and $11\:q_l$ in $D_{2wl}$.
It means that when the moist entropy is based on formulations different from $s(\theta_s)$, the compact feature obtained in (\ref{def_sat_D2l}) for $D_{2w}$ is modified.
It is indeed slightly different from $D_{2wl}$ and the approximation $s_1[(\theta_s)_1]$ of this section, or from $D_{EM}$ and E94's formulation $s^{\ast}(\theta^{\ast})$.
The modifications of $D_{DK}$ are more important  with the DK82's version $c_{pd} \: \ln(\theta_q)$, since typical values for $r_{sw}$ are much larger than those for $r_l \approx q_l$.

The non-saturated and saturated  lapse rates may be written as
\begin{align}
  \Gamma_{1/ns}
  & = \:
     \frac{g}{c_{pd}}
     \;
     \frac{T}{T_v}
  \: ,
\label{def_BVF1_G_ns}
\end{align}
and 
\begin{align}
  \Gamma_{1/sw}
  & = \:
     \frac{g}{c_{pd}}
     \;
     \frac{T}{T_v}
     \;
     \frac{D_{3w}}{D_{2wl}}
  \: ,
\label{def_BVF1_G_sw}
\end{align}
where
\begin{align}
  D_{3w}
  & = \: 1 \: + \:
      \left( 1 + \eta \: r_{sw} \right) \:
      \frac{L_{vap} \: q_{sw}}{R_{d}\:T}
  \: . \label{def_sat_D3l} 
\end{align}
The difference between $D_{1w}$ and $D_{3w}$ is that the virtual temperature $T_v$ in (\ref{def_sat_D1l}) is replaced by the actual one $T$ in (\ref{def_sat_D3l}).

%===============
% Figure (3) :  % Pauluis_S_q_t /  T_rho_S_qt_v0.proc  / T_part2 
%===============
\begin{figure}[hbt]
\centering
 \includegraphics[width=0.65\linewidth,angle=0,clip=true]{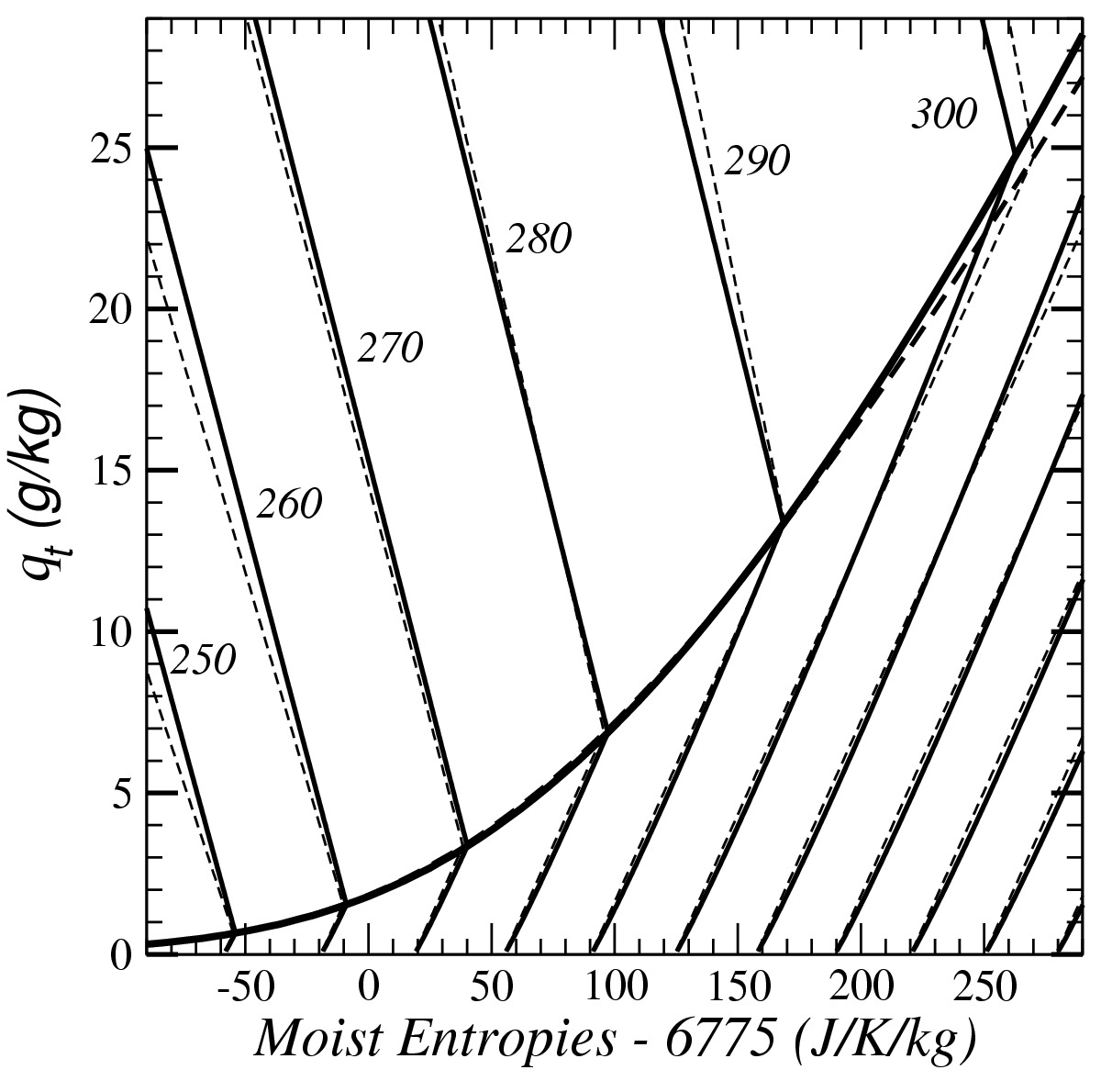}
\caption{\it\small
The same conservative variable diagram as in Figure (\ref{fig_PS10_diagrams}), but with the  exact moist  entropy formulation $s(\theta_s)$ compared to the approximate one $s_1[(\theta_s)_1$] and with extended ranges for the variations of the moist entropies and of $q_t$.
The solid lines correspond to constant values of $T(s, q_t, p_1)$ computed with $s(\theta_s)$ given by  (\ref{def_S_unsat}) and (\ref{def_S_sat}).
The dashed lines correspond to constant values of $T(s_1, q_t, p_1)$ computed with the approximate version $s_1[(\theta_s)_1]$ coming from (\ref{def_S_THs1}) and  (\ref{def_THs1}).
The  entropies are plotted as $s-s^0_{d}$ and $s_1-s^0_{d}$ on the x-axis, with $s^0_{d}=6775$~J~K${}^{-1}$~kg${}^{-1}$.
Lines of equal temperature are labelled every $10$ K.
\label{fig_THS_THS1_diagrams}}
\end{figure}

Comparison of the conservative variable diagrams plotted in Figures~\ref{fig_PS10_diagrams} and \ref{fig_THS_THS1_diagrams} shows that the impacts of the difference between the specific moist entropy formulation $s[(\theta_s)]$ and the approximate version $s_1[(\theta_s)_1]$ are much smaller than the impacts of the use of the Puluis and Schumacher (2010) moist entropy formulation $s^{\ast}(\theta^{\ast})$.
This allows the possible use of $s_1[(\theta_s)_1]$ as an accurate approximation of $s[(\theta_s)]$, whatever the non-saturated or (over-) saturated conditions may be.

As an example of possible application, it is  possible readily to convert the ``bridging relationships'' (\ref{def_FC}) to (\ref{def_BVF_FC_MC}) to the case of $\theta_s \approx (\theta_s)_1$. 
If $c_{pd}$ replaces $c_p$ in the definition of $F(C)$, and if the last term of (\ref{def_sat_D2l1}) appears in the lower case expression of $M(C)$, the equivalent of (\ref{def_BVF_FC_MC}) is then simply expressed by:
\begin{align}
  &N^2_1(C)
  \: = \: g \: M(C)
     \left(
      \frac{\partial \ln(\theta_s)_1 }{\partial z}
     \right)_E
+ \: g\: 
     \left(
      \frac{\partial \ln(q_d)  }{\partial z}
     \right)_E
+ \: g \: M(C)
   \left[ (1+r_v)
    \frac{R_v}{R} \:F(C)
    - \Lambda_r
   \right]_E
     \left(
      \frac{\partial q_t  }{\partial z}
     \right)_E
  .
\label{def_BVF_FC_MC_1}
\end{align}
This more compact formulation allows us better to understand the purpose and limitations of the introduction of $C$ as control (or transition) parameter for the bridging step synthetically described by (\ref{def_BVF_FC_MC_1}).

Potential applications of a formula like (\ref{def_BVF_FC_MC_1}) may correspond (pending other suggestions left to the interested readers) to \begin{itemize}[label=-,leftmargin=3mm,parsep=0cm,itemsep=0cm,topsep=0cm,rightmargin=2mm]
\item the computation of a $N^2$-linked physical quantity like the conversion term of turbulent kinetic energy into other forms of energy;
\item the calculation of a Richardson number (or any 2D or 3D related quantity);
\item some more complex computations, like those of the reduced complexity model for interactions between moist convection and gravity waves described in Ruprecht et al. (2010) and Ruprecht and Klein (2011).
There, $C$ would represent de facto the area fraction of cloudy air in horizontal slices.
\end{itemize}

In all cases, the question of the definition (or the parametrization) of the control parameter $C$ becomes a central one and this issue is likely to take a differing shape from case to case.
If seeking full complexity, $C$ should not be confused with a proportion of saturated air within the considered air parcel.
There are two reasons for this.
First, as explained in Appendix~F, there is no reason to consider $N^2(C)$ or $N^2_1(C)$ as a $C$-weighted linear interpolation between the extreme cases of fully unsaturated and fully saturated conditions. 
Second, in most conditions, there would exist (partly) organized motions differentiating the mean dynamical behaviour of  the clear-air and cloudy patches of the considered air parcel, respectively.
Nonetheless, we may postulate a monotonic dependency of $C$ on the above-mentioned proportion.

Despite the weakness linked to the generally heuristic character of the definition of $C$, two additional remarks support the potential use of (\ref{def_BVF_FC_MC}) or (\ref{def_BVF_FC_MC_1}).

First, in the already mentioned case of FIRE-I marine Stratocumulus clouds, there is hardly any gradient of specific moist entropy between clear-air and cloudy patches in Figure~\ref{Fig_RF03B_N2_results}~(a), and most of the so-called subgrid transport of $q_t$ is ensured by turbulent motions and not by partly organized compensating motions between these patches. 
Hence, viewing here $C$ as a kind of subgrid cloud cover becomes rather legitimate, if one indeed attributes the nonlinear part of the $N^2(C)$ or $N^2_1(C)$ behaviour to the existence of the above-mentioned (small) partly organized transport.

Second, the fact that one single parameter is sufficient to obtain a monotonic, general and consistent transition between unsaturated and saturated situations is a welcome step for applications seeking a robust and simple behaviour. 

In summary, the proposed transition formulas need to be used with a lot of care (especially for the estimation of $C$): there is at least one case where they should be directly appropriate, while in other cases they may be useful because of their simplicity (one control parameter alone) and of their monotonic character, for lack of other alternatives in front of practical problems.

%==================================
\section{Some numerical applications.}    %  (Section  9)
%==================================
\label{sec_N2m_Appl_Num}

A numerical application is presented in this section by using the same RF03B FIRE-I observations as in M11, except with the additional constraint that $q_v \equiv q_{sw}$ if $q_l > 0.1$~g/kg and that $q_l \equiv 0$ if $q_l < 0.1$~g/kg.
The profiles have been slightly filtered vertically, in order to give smoother profiles and less noisy vertical gradients.
The same average profiles are used for all the formulations of $N^2$.

\begin{figure}[hbt]
\centering
(a)\includegraphics[width=0.31\linewidth,angle=0,clip=true]{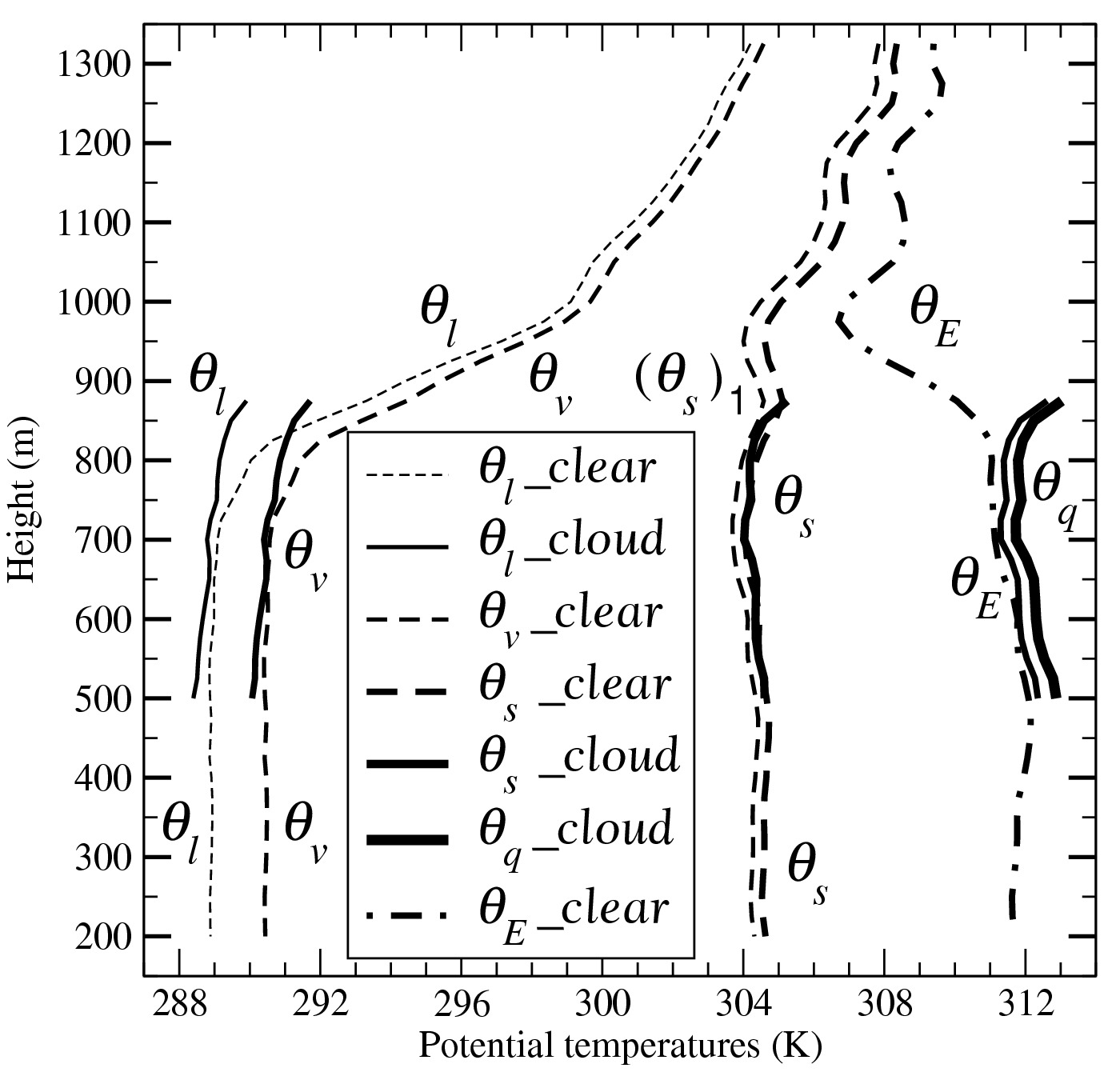}
(b)\includegraphics[width=0.23\linewidth,angle=0,clip=true]{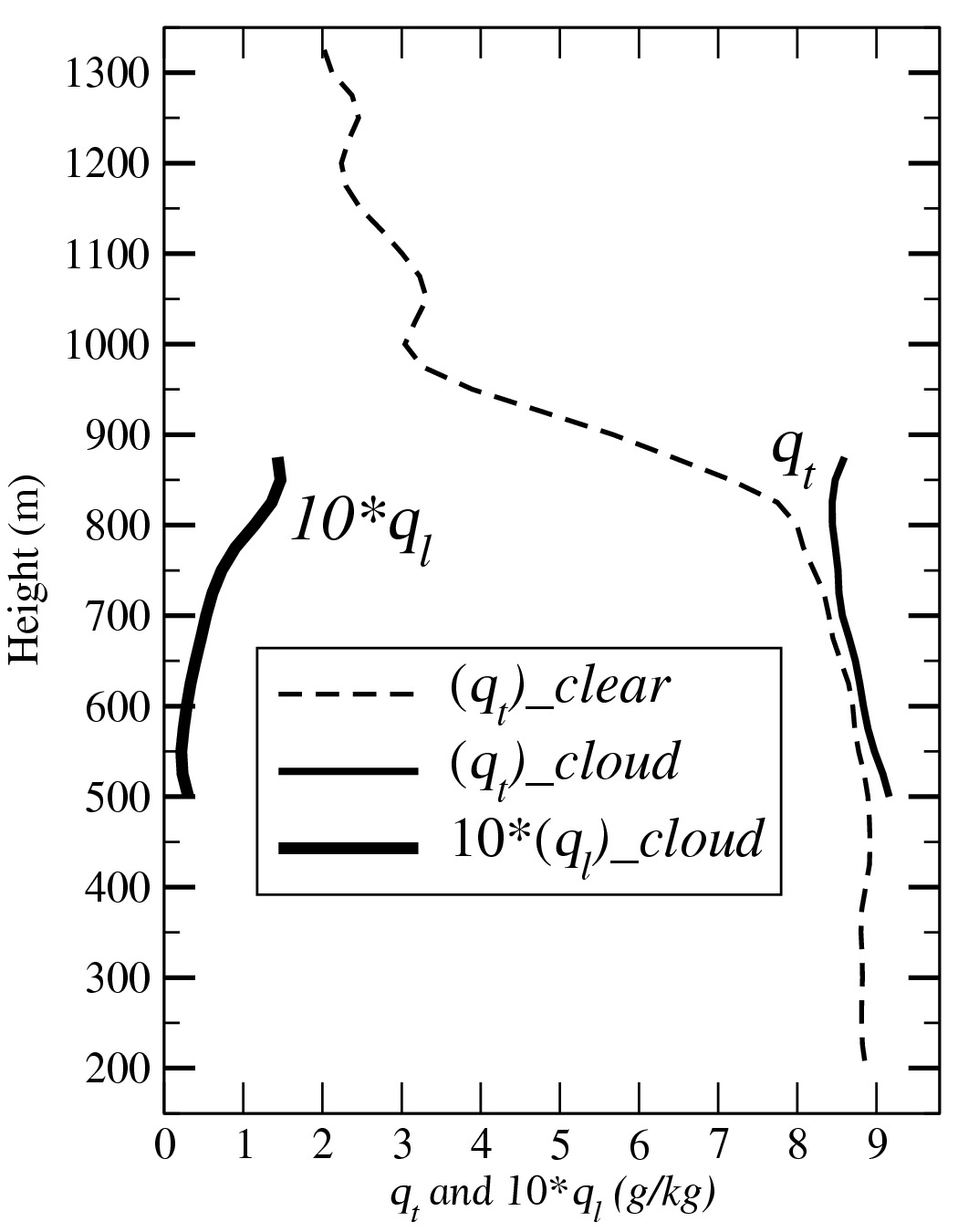}
(c)\includegraphics[width=0.23\linewidth,angle=0,clip=true]{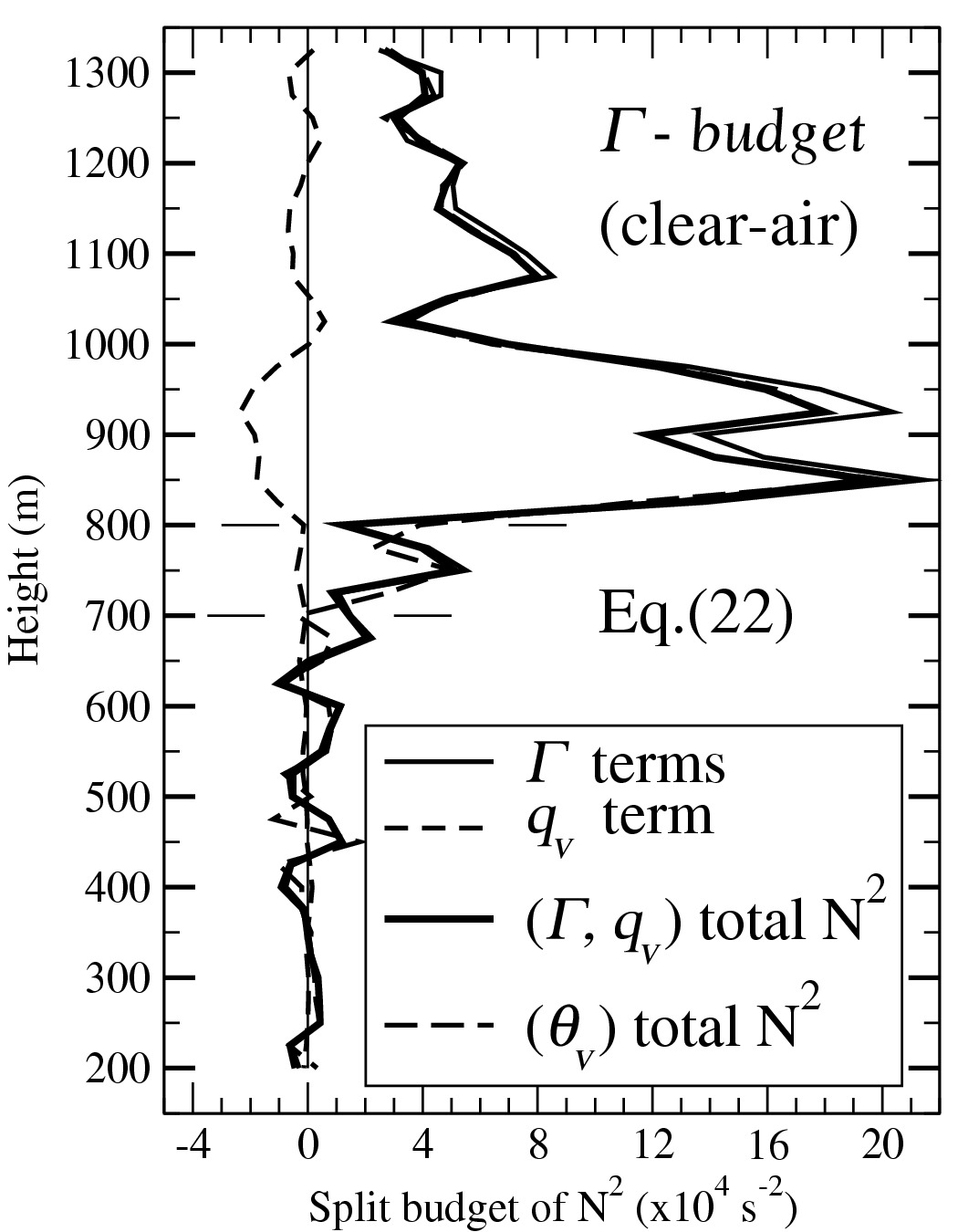}
\\
(d)\includegraphics[width=0.23\linewidth,angle=0,clip=true]{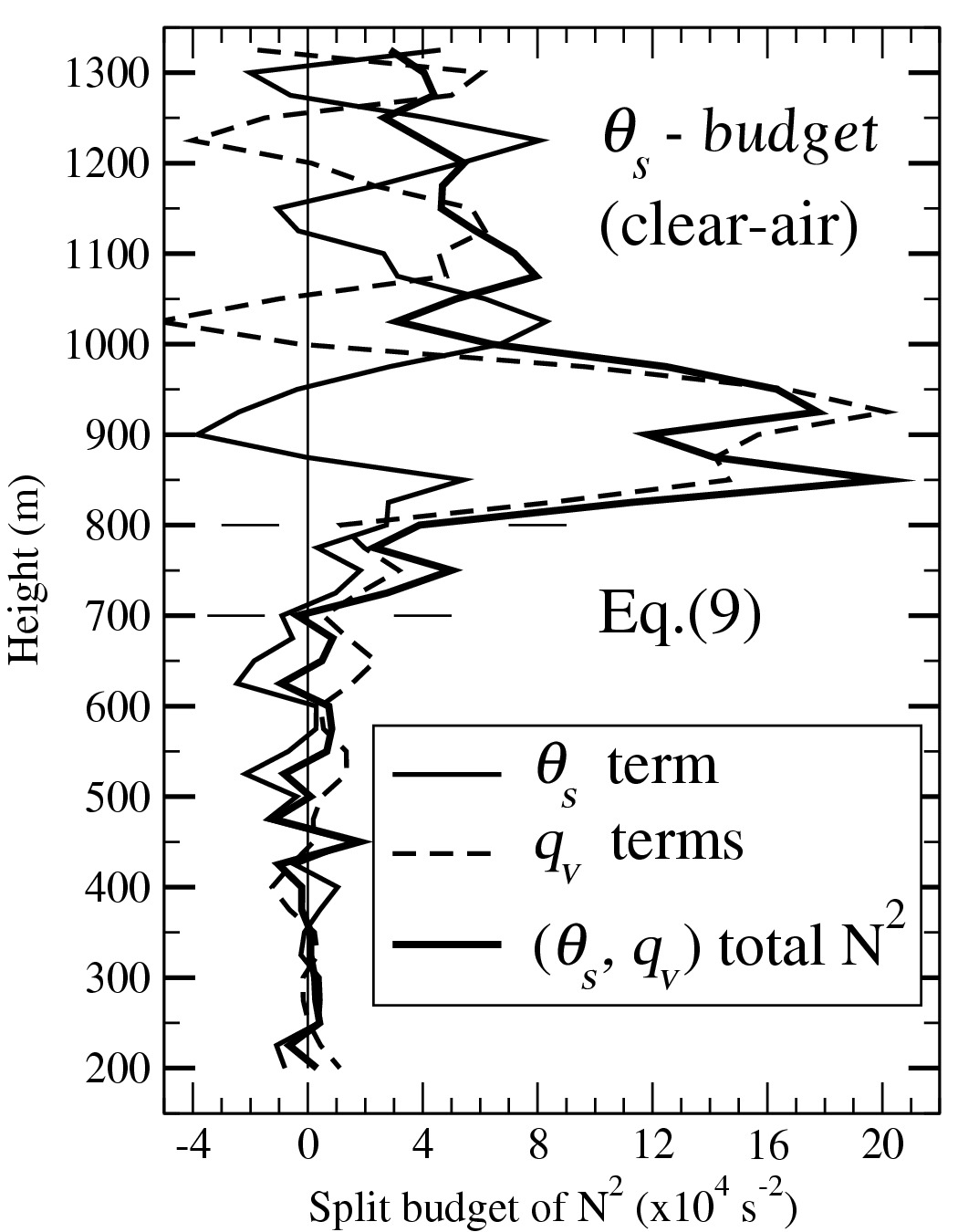}
(e)\includegraphics[width=0.23\linewidth,angle=0,clip=true]{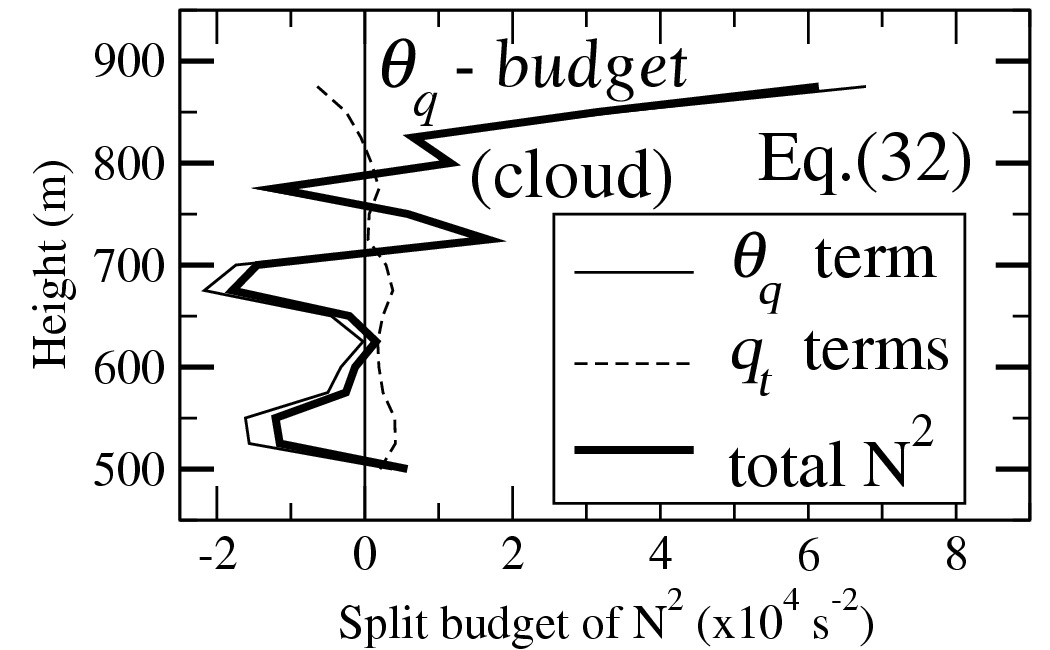}
(f)\includegraphics[width=0.23\linewidth,angle=0,clip=true]{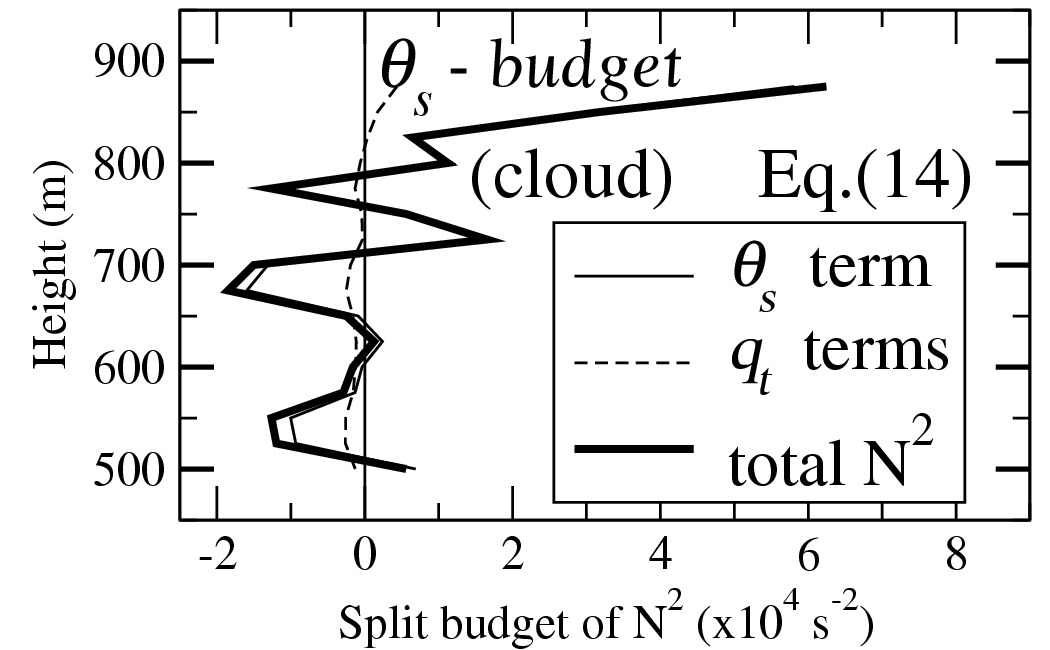}
\caption{\it\small
A study of the FIRE-I (RF03B) Stratocumulus (see M11).
The vertical profiles of ${\theta}_l$, ${\theta}_v$, $({\theta}_s)_1$,  ${\theta}_s$,  ${\theta}_E$ and  ${\theta}_q$ are depicted in (a), from left to right.
The vertical profiles of $q_t$ and $10 \times q_l$ are depicted in (b).
On both (a) and (b) the solid (dashed) lines represent in-cloud (clear-air) and saturated (unsaturated) conditions.
Only in-cloud values of ${\theta}_q$ are plotted, since the formulation (\ref{def_Theta_q}) is not valid for  clear-air (unsaturated) conditions.
The vertical profile of the clear-air approximate formulation $(\theta_s)_1$ is located close to the clear-air profile for $\theta_s$, with an almost constant bias of about $-0.4$ to $-0.6$~K (see M11).
Units are in K for the potential temperatures and in g/kg for the water contents.
The budgets of the moist squared BVF equations are depicted in (c)-(f).
The total budgets (heavy solid lines) are split into the terms depending either on the lapse-rate or gradients of potential temperatures (solid lines) or on the gradients of the water contents $q_v$ or $q_t$ (dashed lines).
The usual Lapse-rate budget (\ref{def_BVF_ns_for_DK82}) of the clear-air (unsaturated) squared BVF formulation is depicted in the panel (c).
The standard formulation (\ref{def_BVF_N2v_ns}) for $N^2_v$ expressed in terms of the gradient of $\theta_v$ is depicted bu the long-dashed line.   
The new clear-air budget (\ref{def_BVF_ns_new}) using the (unsaturated) $\theta_s$ specific moist entropy formulation is depicted in the panel (d).
The DK82 squared BVF formulation (\ref{def_BVF_sat_DK82_bis}) valid for in-cloud (saturated) conditions and using $\theta_q$ is depicted in (e).
The new in-cloud (saturated) $\theta_s$ specific moist entropy formulation (\ref{def_BVF_sat_new_l}) is depicted in (f).
All the squared BVF values are multiplied by a factor of $10^4$, with units in $s^{-2}$.
\label{Fig_RF03B_N2_results}}
\end{figure}

The vertical profiles of the basic variables are depicted in the Figs.(\ref{Fig_RF03B_N2_results}) (a) and (b), where the $\theta_q$ curve given by (\ref{def_Theta_q}) appears to be similar to the saturated $\theta_E$ one, with $\theta_s \approx (\theta_s)_1 \approx 304.5$~K in a two-third position between $\theta_l \approx \theta_v \approx 288$~K and $\theta_E \approx \theta_q \approx 312$~K, as already indicated in the Fig.(11-b) of M11.

%===============
% Figure (4) :  : sous FIRE_I / zprog_calcul_9b_N2MG_review
%===============
%- - - - - - - - - - - - - - - - - - - - - - - - - -
% 4-a  THl_THs   : ITRACE_THl_THs   ; ICONVERT0
% 4-b  Q_tot_liq : ITRACE_Qt_Ql     ; ICONVERT1
% 4-c  Tv_clr    ; ICONVERT1
% 4-d  THs_clr   ; ICONVERT1
% 4-e  THq_cld   : ITRACE_N2_THq_Qt ; ICONVERT2
% 4-f  THs_cld   : ITRACE_N2_THs_Qt ; ICONVERT2
%- - - - - - - - - - - - - - - - - - - - - - - - - -

The clear-air profiles of the Betts' variable $\theta_l$ and $q_t$ are clearly different from the associated in-cloud profiles in the upper-PBL entrainment region.
The same feature is valid for the $\theta_v$ and $\theta_q$ curves (the two profiles used in DK82).
In contrast, the clear-air and in-cloud vertical profiles of $\theta_s$ are almost superimposed, illustrating the full mixing in specific moist entropy within the stratocumulus, as already mentioned in M11.

Moreover, the $\theta_q$ and $\theta_E$ profiles are almost similar but far from the $\theta_v$ curve, with almost opposite vertical gradients.
This must correspond to a less continuous feature between the clear-air and in-cloud formulations of $N^2$ with the standard DK82 approach, where the non-saturated budget of $N^2$ is based on $\theta_v$ and the saturated one is expressed in terms of $\theta_q \approx \theta_E$.

Let us comment the Figures 4 (c)-(f).
For the standard clear-air formulation (\ref{def_BVF_ns_for_DK82}) which uses the lapse-rate approach, the budget of the squared BVF is dominated in (c) by the thermal component, with a much smaller water content component.
For the new formulation in (d), which uses the conservative variables ($\theta_s, q_t$) approach, the (total) value of $N^2$ is almost the same as for DK82, but the clear-air budgets is made of large and compensating thermal versus water content components (within the whole moist PBL and the dry air above as well).

A more detailed analysis shows that some numerical differences exist between $650$ and $800$~m (thin lines have been added at $700$ and $800$~m), where the total $N^2$ budgets (thick solid lines) are not the same in (c) and in (d).
It appears that the vertical profile of $T$ (not shown) exhibits more noisy and uneven vertical shape than the vertical profiles of $q_v$, leading to gradients of $\theta_s$ which are easier to determine than the lapse rates.
Since the standard formulation (\ref{def_BVF_N2v_ns}) for $N^2_v$ -- long-dashed line in (c) -- is very close to the ($\theta_s, q_t$) approach -- thick line in (d) --, it may be concluded that the differences are  due to less accurate evaluations of the stability feature  using the lapse rate method than with the other methods based on virtual or specific moist entropy potential temperatures.

The in-cloud budgets presented in (e) and (f) show that the total values of saturated $N^2$ are almost the same for the  DK82 formulation with $\ln(\theta_q)$ as for the new formulations with $\ln(\theta_s)$.
However, for the new formulation in (f), the water content component is of opposite sign and is more ``neutral'' than in (e), i.e. it is closer to $0$ at each level.
These differences are the consequences of the second line in (\ref{def_BVF_sat_new_l}) which does not exists in (\ref{def_BVF_sat_DK82_bis}), leading to a different partitioning of the budgets of $N^2$.

% ------------------------------------------------
\section{Conclusions.} % (section 10)
% ------------------------------------------------
\label{section_CONCLUDE}

Both non-saturated and saturated versions of the moist squared BVF ($N^2$) have been computed in terms of the vertical gradients of the moist natural conservative variables, namely the specific content of dry air (or total water)  and the specific moist entropy.
The latter has been defined in terms of the specific entropy potential temperature (\ref{def_THs})  for $\theta_s$ introduced in M11, differently from the moist entropies and potential temperatures already defined in DK82, E94, P08 or P11.

Comparisons with the previous results published in DK82 and E94 show that the adiabatic lapse rates are different.
The conservative variable diagrams published in P11 are also modified if the present $\theta_s$ formulation for the specific moist entropy is used, with possible different physical properties.
A new small counter-gradient term appears when the new non-saturated version of $N^2$ is written in terms of the vertical gradient of the virtual potential temperature.

Numerical applications made with the FIRE-I data sets indicate that there is little difference for the (total) values of $N^2$.
Larger impacts are observed if the budget of $N^2$ is  partitioned into a sum of separated terms depending on gradients of $s$ and $q_d$ (first lines) and $q_t$ (second lines), with weighting factors really different from the ones obtained with DK82 or E94 moist entropies.

It is possible to replace $\theta_s$  by the approximate version $(\theta_s)_1$ and to derive a corresponding approximate formulations for  $N^2$.
A continuous transition is suggested between the new non-saturated and saturated versions of $N^2$, leading to possible definition of a control parameter $C$ valid for both $(\theta_s)_1$ and $\theta_s$ formulations.

It is a kind of paradox that the complexity of the specific moist entropy defined with the full formulation of $\theta_s$ should lead to rather simple and compact formulations for $N^2_{ns}$ and $N^2_{sw}$.
In particular, the terms $D_{DK}$ and $D_{EM}$ involved in the DK82's and E94's computations of the moist, saturated, adiabatic lapse rate and the associated squared BVFs are more complicated than with the formulations $\Gamma_{sw}$ and $N^2_{sw}$ based on $\theta_s$.
Additional small terms depending on the liquid-water content appear in $D_{DK}$ and $D_{EM}$, and they disappear in $D_{sw}$.

An explanation for this paradox could be found in the complex moist basic formulas like (\ref{def_prop_sat}) and (\ref{def_Lv_0}) which define $c_p$ or $L_{vap}(T)$, among others.
They both depend on the thermodynamic properties of water species in such a way that the full specific entropy formulation (\ref{def_THs}) for $\theta_s$ seems to be required in order to arrive at a cancellation of all small terms.
If approximations are made in the definitions of the moist entropies, like the hypothesis of zero dry-air and liquid-water reference entropies in E94, or in the definitions of $R$, $c_p$ or $L_{vap}$, the cancellation of the small terms is incomplete.
It is true, for instance, with $\theta_q$ and $\theta^{\star}$ used as starting points to compute $N^2$.

It may be worthwhile to note that the approximation of $\theta_s$ by $(\theta_s)_1$ generates different but still simple and compact versions of $N^2_{1/ns}$ and $N^2_{1/sw}$.
As for the DK82's and E94's versions, the corresponding term $D_{2wl}$ contains an additional small term depending on $q_l$, but the same continuous transition with the same parameter $C$ is obtained between $N^2_{1/ns}$ and $N^2_{1/sw}$ as between $N^2_{ns}$ and $N^2_{sw}$.
In this respect, it may indicate that the approximation of $\theta_s$ by $(\theta_s)_1$ is of smaller impact than the use of $\theta_q$ or $\theta^{\star}$.

%-------------------------------------------------------------------------
%    Acknowledgement(s) : \acknowledgement or \acknowledgements
%-------------------------------------------------------------------------

%%%%%%%%%%%%%%%%%%%%
% ACKNOWLEDGEMENTS %
%%%%%%%%%%%%%%%%%%%%%%%%%%%%%%%%%%%%%%%%%%%%%%%%%%%%%%%%
% The title "Acknowledgement"  is provided by : \ack   %
% The title "Acknowledgements" is provided by : \acks  %
%%%%%%%%%%%%%%%%%%%%%%%%%%%%%%%%%%%%%%%%%%%%%%%%%%%%%%%%

\vspace{5mm}
\noindent{\large\bf Acknowledgements}
\vspace{2mm}

The authors are most grateful to D. Mironov for initially insisting on the potential importance of this work as well as for further exchanges, to R. De Troch for stimulating discussions and to O. Pauluis for his encouragements to go forward with the complex analytical task. 
Both Rupert Klein and the other anonymous reviewer have clearly helped to improve the scope and content of the article by requiring a more general and more ambitious approach, something eventually much appreciated by the authors. 
Most of this work was completed in the spirit and framework of the EU-funded COST ES0905 action.

The validation data from the NASA Flights during the FIRE I experiment have been kindly provided by S. R. de Roode and Q. Wang.

%-----------------
%   APPENDIXES
%-----------------

%----------------- -------------------------------------
%    APPENDIX - A    
%----------------- -------------------------------------
%----------------------------------------------------------------------
\vspace{4mm}
\noindent
{\bf Appendix A. List of symbols and acronyms.}
%----------------------------------------------------------------------
             \label{appendixSymbol}
\renewcommand{\theequation}{A.\arabic{equation}}
  \renewcommand{\thefigure}{A.\arabic{figure}}
   \renewcommand{\thetable}{A.\arabic{table}}
      \setcounter{equation}{0}
        \setcounter{figure}{0}
         \setcounter{table}{0}
\vspace{1mm}
\hrule

\begin{tabbing}
--------------\=  -------------------------------------- --\= \kill
BVF    \> Brunt-V\"{a}is\"{a}l\"{a} Frequency \\
FIRE   \>  the First ISCCP Regional Experiment\\
ISCCP  \> International Satellite Cloud Climatology Project \\
PBL    \> Planetary Boundary Layer \\
$(\partial/\partial z)_{par}$ \> Gradients computed for the parcel  \\
$(\partial/\partial z)_{env}$ \> Gradients computed for the environment  \\
$C$, $C_0$ \> control parameters in the Appendix~F\\
$c_{pd}$ \> specific heat for dry air   \>($1004.7$~J~K${}^{-1}$~kg${}^{-1}$) \\
$c_{pv}$ \> spec. heat for water vapour \>($1846.1$~J~K${}^{-1}$~kg${}^{-1}$) \\
$c_{l}$  \> spec. heat for liquid water \>($4218$~J~K${}^{-1}$~kg${}^{-1}$) \\
$c_{i}$  \> spec. heat for ice          \>($2106$~J~K${}^{-1}$~kg${}^{-1}$) \\
$c_p$ \> specific heat at constant pressure for moist air, \\
      \> $ = \: q_d \: c_{pd} + q_v \: c_{pv} + q_l \: c_l  $
         $ = \: q_d \: ( \: c_{pd} + r_v \: c_{pv} + r_l \: c_ l)$ \\
$c_p^{\ast}$ \> $ = \: c_{pd} + r_{sw} \: c_{pv} $ (E94's formulation)  \\
$D_{1w}$ \> a shortcut notation, like $D_{2w}$, $D_{DK}$, $D_{EM}$, ...  \\
$\delta$ \> $=R_v/R_d-1 \approx 0.608$ \\
$\eta$  \> $=1+\delta =R_v/R_d \approx 1.608$ \\
$\varepsilon$ \> $=1/\eta=R_d/R_v \approx 0.622$ \\
$\kappa$ \> $=R_d/c_{pd}\approx 0.2857$ \\
$\gamma$ \> $= \eta \: \kappa \ = R_v/c_{pd} \approx 0.46$ \\
$\lambda$ \> $= c_{pv}/c_{pd}-1 \approx 0.8375$ \\
$e$      \> the water-vapour partial pressure \\
$e_{sw}(T)$ \> partial saturating pressure over liquid water \\
$e_r$      \> the water vapour reference partial pressure:
             $\: e_r = e_{ws}(T_0) \approx 6.11$~hPa \\
$F(C)$ \> a function of $C$, like $M(C)$  \\
$g$      \> Gravity's constant ($9.80665$~m~s${}^{-2}$) \\
$\Gamma_{ns}$    \> the  lapse rate ($-\partial T/\partial z$ / unsaturated)  \\
$\Gamma_{sw}$    \> the liquid-water saturated version of $\Gamma_{ns}$ \\
$\Lambda_r$ \> $= [ (s_{v})_r - (s_{d})_r ] / c_{pd} \approx 5.87$ \\
$\Lambda_{v}$  \> an additional  term to $\Lambda_r$ ($\Lambda_{sw}$ as well) \\
$L_{vap} (T)$ \> $=h_v-h_l$: latent heat of vaporisation \\
$L_{vap} (T_0)$ \> $= 2.501$~$10^{6}$~J~kg${}^{-1}$ \\
$L_{sub} (T)$ \> $=h_v-h_i$: latent heat of sublimation \\
$L_{sub} (T_0)$ \> $= 2.835$~$10^{6}$~J~kg${}^{-1}$ \\
$L_v^0$ \> a latent heat shortcut notation  \\
$N^2$  \> squared BVF notations ($N^2_m$, $N^2_{ns}$, $N^2_{DK}$, ...) \\
$p$      \> $=p_d + e$: local value for the pressure \\
$p_r$  \> $=(p_d)_r + e_r$: reference pressure ($p_r=p_0$)\\
$p_d$    \> local dry-air partial pressure \\
$(p_d)_r$ \> reference dry air partial pressure ($\equiv p_r-e_r$)\\
$p_0$    \> $=1000$~hPa: conventional pressure \\
$\psi$  \> a dummy variable (section~7 and Appendix~B) \\
$q_{d}$  \> $={\rho}_d / {\rho}$: specific content for dry air \\
$q_{v}$  \> $={\rho}_v / {\rho}$: specific content for water vapour \\
$q_{l}$  \> $={\rho}_l / {\rho}$: specific content for liquid water \\
$q_{i}$  \> $={\rho}_li/ {\rho}$: specific content for solid water \\
$q_{sw}$  \> specific content for saturating water vapour \\
$q_t  $  \> $= q_v+q_l+q_i$: total specific content of water \\
$r_{v}$  \> $=q_{v}/q_{d}$: mixing ratio for water vapour \\
$r_{l}$  \> $=q_{l}/q_{d}$: mixing ratio for liquid water \\
$r_{i}$  \> $=q_{i}/q_{d}$: mixing ratio for solid water \\
$r_{r}$  \> reference mixing ratio for water species: 
                 $\eta\:r_{r} \equiv e_r / (p_d)_r$
             and $r_{r} \approx 3.82$~g~kg${}^{-1}$ \\
$r_{sw}$  \> mixing ratio for saturating water vapour \\
$r_{t}$  \> $=q_{t}/q_{d}$: mixing ratio for total water \\
${\rho}_d$  \> specific mass for the dry air  \\
${\rho}_v$  \> specific mass for the water vapour \\
${\rho}_l$  \> specific mass for the liquid water \\
${\rho}_i$  \> specific mass for the solid water \\
${\rho}$  \> specific mass for the moist air $={\rho}_d+{\rho}_v+{\rho}_l+{\rho}_i$  \\
$R_v$  \> water vapour gas constant --\= ($461.52$~J~K${}^{-1}$~kg${}^{-1}$) \kill
$R_d$  \> dry air gas constant    \> ($287.06$~J~K${}^{-1}$~kg${}^{-1}$) \\
$R_v$  \> water vapour gas constant \> ($461.53$~J~K${}^{-1}$~kg${}^{-1}$) \\
$R$    \> $ = q_d \: R_d + q_v \: R_v$: gas constant for moist air 
          $ = q_d \:(\: R_d + r_v \: R_v)$ \\
$s$      \> the specific moist entropy associated with ${\theta}_{s}$ \\
$s_1$  \> the specific moist entropy associated with $({\theta}_{s})_1$ \\
${s}_{ref}$  \> a reference specific entropy  \\
${s}_d$  \> specific entropy for the dry air  \\
${s}_v$  \> specific entropy for the water vapour \\
${s}_l$  \> specific entropy for the liquid water \\
$s^{\ast}$  \> a moist entropy (E94) \\
$(s_{d})_r$  \>  reference values for the entropy of dry air
                 at $T_r$ and $(p_d)_r$ \\
$(s_{v})_r$  \> reference values for the entropy of water vapour
                at $T_r$ and $e_r$\\
$s^0_d$  \> standard specific entropy for the dry air
            at $T_0$ and $p_0$: $6775$~J~K${}^{-1}$~kg${}^{-1}$) \\
$s^0_v$  \> standard specific entropy for the water vapour
            at $T_0$ and $p_0$: $10320$~J~K${}^{-1}$~kg${}^{-1}$) \\
$T$      \> local temperature \\
${T}_{v}$  \> virtual temperature associated to  ${\theta}_{v}$  \\
${T}_{\rho}$  \> E94's version for  ${T}_{v}$  \\
$T_{r}$  \> the reference temperature ($T_r\equiv T_0$) \\
$T_{0}$  \> zero Celsius temperature ($=273.15$~K) \\
$\theta$        \> $ = T\:(p_0/p)^{\kappa}$: potential temperature\\
${\theta}^{\ast}$  \> a moist entropy potential temperature (E94) \\
${\theta}_{q}$  \> a  moist entropy potential temperature (DK82) \\
${\theta}_E$  \> equivalent potential temperature \\
${\theta}_{v}$  \> virtual potential temperature  \\
${\theta}_{l}$  \> liquid-water potential temperature \\
${\theta}_{s}$  \> specific moist entropy potential temperature (M11) \\
$({\theta}_{s})_1$  \> approximate version of ${\theta}_{s}$ \\
$z$  \> vertical coordinate
\end{tabbing}

\vspace{1mm}
\hrule

%----------------- -------------------------------------
%    APPENDIX - B   
%----------------- -------------------------------------
%----------------------------------------------------------------------
\vspace{4mm}
\noindent
{\bf Appendix B. General squared BVF formulations.}
%----------------------------------------------------------------------
\label{appendix_N2_Moist}
\renewcommand{\theequation}{B.\arabic{equation}}
  \renewcommand{\thefigure}{B.\arabic{figure}}
  \renewcommand{\thetable}{B.\arabic{table}}
      \setcounter{equation}{0}
        \setcounter{figure}{0}
        \setcounter{table}{0}
\vspace{1mm}

The method for computing the moist value of $N^2$ is usually based on the classical approach of DK82, where the adiabatic changes of the density of the parcel are evaluated and compared to the corresponding values of the environment, leading to
\begin{align}
  N^2_{m}  & = \:
      \left(
        \frac{g}{{\rho}}\:
        \frac{\partial{\rho}}{\partial z}
      \right)_{par.}
  \; - \;
      \left(
        \frac{g}{{\rho}}\:
        \frac{\partial{\rho}}{\partial z}
      \right)_{env.}
  \: . \label{def_BVF0}
\end{align}

Let us consider the method based on the material published in PA08 and  PCK10, where it is stated that any thermodynamic variable $\psi$  can be expressed in terms of the entropy $s$, the total water content $q_t$ and the pressure $p$ alone.
It is true in particular for $\psi$  representing any of the temperature ($T$), the specific volume ($\alpha$), the density ($\rho$), the water contents ($q_v$, $q_l$, $q_i$) or the buoyancy ($B$), leading to $\psi(s, q_t, p)$.

It is indeed possible to use the set of three independent variables $(s, q_t, p)$ if it is assumed that a parcel of moist atmosphere is either saturated (with $q_v$ equal to its saturated value and with existing condensed water equal to $q_t - q_v$) or non-saturated (with no condensed water and $q_t = q_v$).
In this way, the condensed water contents $q_l$ and $q_i$ no longer appear as independent variables of the system and they must be derived from the information given by $(s, q_t, p)$, with either liquid water for $T>T_0$ or solid water for $T<T_0$.

The property (\ref{def_new_N2_moist_6}) can be derived through a short mathematical method, starting with (\ref{def_BVF0}) rewritten as
\begin{align}
  N^2_{m}  & = \:
      \left.
        \frac{g}{{\rho}}\:
        \frac{\partial{\rho}}{\partial z}
      \right|_{s,q_t}
  \; - \;
        \frac{g}{{\rho}}\:
        \frac{\partial{\rho}}{\partial z}
  \: . \label{def_BVF1}
\end{align}
From the chain rule, the gradient of the density $\rho(s, q_t, p)$ is equal to
\begin{align}
 \frac{\partial \rho}{\partial z}
  & = \:      \left.
                  \frac{\partial \rho}{\partial s}
                  \right|_{p,q_t}
                \frac{\partial  s  }{\partial z  }
                \:+
                  \left.
                  \frac{\partial \rho}{\partial q_t}
                  \right|_{p,s}
                \frac{\partial q_t }{\partial z  }
                \:+
                  \left.
                  \frac{\partial \rho}{\partial p}
                  \right|_{s,q_t}
                \frac{\partial p }{\partial z  }
  \: . \label{def_BVF2}
\end{align}
If hydrostatic conditions prevail, then $dp = - \rho \: g \: dz$ is applied twice in the last term of (\ref{def_BVF2}), this last term being equal to
\begin{align}
                   \left.
                  \frac{\partial \rho}{\partial z}
                  \right|_{s,q_t}
  & = \:  
                \frac{\partial \rho}{\partial z} 
                 \:-
                \left.
                  \frac{\partial \rho}{\partial s}
                  \right|_{p,q_t}
                \frac{\partial  s  }{\partial z  }
                \:-
                  \left.
                  \frac{\partial \rho}{\partial q_t}
                  \right|_{p,s}
                \frac{\partial q_t }{\partial z  }
  \: . \label{def_BVF3}
\end{align}
The property (\ref{def_new_N2_moist_6}) is obtained with (\ref{def_BVF3}) inserted into (\ref{def_BVF1}).

%----------------- -------------------------------------
%    APPENDIX - C   
%----------------- -------------------------------------
%----------------------------------------------------------------------
\vspace{4mm}
\noindent
{\bf Appendix C. The unsaturated moist squared BVF.}
%----------------------------------------------------------------------
\label{appendix_unsat_BVF}
\renewcommand{\theequation}{C.\arabic{equation}}
  \renewcommand{\thefigure}{C.\arabic{figure}}
  \renewcommand{\thetable}{C.\arabic{table}}
      \setcounter{equation}{0}
        \setcounter{figure}{0}
        \setcounter{table}{0}
\vspace{1mm}

The properties $q_l=q_i=0$ and $q_t=q_v$ are used to derive the unsaturated moist air version of the state equation (\ref{def_eq_state}) and of the virtual temperature definition (\ref{def_Tv2}), resulting in
\begin{align}
  \rho (T, q_t, p)  &  \: = \;
  \frac{1}{T} \;
  \frac{1}{1+\delta\:q_t} \;
  \frac{p}{R_d}
  \: . \label{def_rho_unsat}
\end{align}

The properties $q_l=q_i=0$, $q_t=q_v$ and $r_t=r_v$ are used to transform the definitions (\ref{def_S_THs}) and (\ref{def_THs}) to express the entropy for unsaturated moist air as
\begin{align}
  s (T,q_t,p)
& \: = \; s_{ref}
  + {c}_{pd} \: \ln\left( T \right)
  + {c}_{pd} \: \lambda \: q_t \: \ln\left( \frac{T}{T_r} \right)
\nonumber \\
& \quad
  - R_d \: \ln\left( \frac{p}{p_0} \right)
  + {c}_{pd} \: \Lambda_r \: q_t
  - {c}_{pd} \: \kappa \: \delta \: q_t \: \ln\left( \frac{p}{p_r} \right)
\nonumber \\
& \quad
  - {c}_{pd} \: \gamma \: q_t \: \ln\left( \frac{r_t}{r_r} \right)
  + {c}_{pd} \: \kappa \: \delta \: q_t \:
                \ln\left( \frac{1+\eta\:r_t}
                                {1+\eta\:r_r}
                    \right)
  + {c}_{pd} \: \kappa \:
                \ln\left( 1+\eta\:r_t \right)
    . \,
  \label{def_S_unsat}
\end{align}

The first partial derivative of $\rho$ with respect to $s$ (at constant values for $p$ and $q_t$)  is computed from  (\ref{def_rho_unsat}) and with $T=T(s,q_t,p)$, leading to
\begin{align}
  \left.
  \frac{\partial \rho}{\partial s}
  \right|_{p,q_t}
  & = \: - \:
  \frac{\rho}{T} \;
  \left.
  \frac{\partial T}{\partial s}
  \right|_{p,q_t}
  \: . \label{def_ns_drho_ds}
\end{align}

The partial derivative of $T$ with respect to $s$ (at constant values for $p$ and $q_t$)  is obtained by computing the derivative of (\ref{def_S_unsat}) with respect to $s$ and with $T=T(s,q_t,p)$ (i.e. involving only the two terms in the first line on the R.H.S.), leading to
\begin{align}
  \left.
  \frac{\partial s}{\partial s}
  \right|_{p,q_t}
  & \: \equiv \; 1 \: = \:
  {c}_{pd} \: \left( \frac{1+\lambda\:q_t}{T} \right) \;
  \left.
  \frac{\partial T}{\partial s}
  \right|_{p,q_t}
  \: , \label{def_ns_dT_ds_1} \\
  \frac{1}{T} \:
  \left.
  \frac{\partial T}{\partial s}
  \right|_{p,q_t}
  & \: = \: \frac{1}{c_{pd}\:(1+\lambda\:q_t)}
    \: = \: \frac{1}{c_p}
  \: . \label{def_ns_dT_ds_2}
\end{align}
From (\ref{def_ns_drho_ds}) and (\ref{def_ns_dT_ds_2}), the first partial derivative involved in the  formulation  (\ref{def_new_N2_moist_6}) for the non-saturated version of $N_m^2$ is  equal to
\begin{equation}
  \left.
  \frac{\partial \rho}{\partial s}
  \right|_{p,q_t}
  \: = \: - \: \frac{\rho}{c_p}
  \: . 
  \label{def_ns_drho_ds_2}
\end{equation}

The second partial derivative of $\rho$ with respect to $q_t$ (at constant values for $p$ and $s$) is computed from (\ref{def_rho_unsat}) and with $T=T(s,q_t,p)$, leading to
\begin{align}
  \left.
  \frac{\partial \rho}{\partial q_t}
  \right|_{p,s}
  & = \: - \:
  \frac{\rho}{T} \;
  \left.
  \frac{\partial T}{\partial q_t}
  \right|_{p,s}
\: - \:
  \rho
  \left(
  \frac{R_v}{R} - \frac{T}{T_v}
  \right)
  \: . \label{def_ns_drho_dqt}
\end{align}

The partial derivative of $T$ with respect to $q_t$ (at constant values for $p$ and $s$) is obtained by computing the derivative of (\ref{def_S_unsat}) with respect to $q_t$ and with $T=T(s,q_t,p)$, to arrive at
\begin{align}
  \left.
  \frac{\partial s}{\partial q_t}
  \right|_{p,s}
  & \equiv \; 0 \: = \:
  \frac{c_p}{T}
  \left.
  \frac{\partial T}{\partial q_t}
  \right|_{p,s}
  + \:
  c_{pd} \:
  \left(\:
  \Lambda_r + \Lambda_v
  \:\right)
  \: , \label{def_ns_dT_dqt_1}
\end{align}
where $\Lambda_v$ is given by (\ref{def_ns_Lambda_v}).
This term is equal to $0$ for the reference conditions $T=T_r$, $p=p_r$ and $r_v=r_r$ and, as such, it is expected to be a corrective term to $\Lambda_r$.

It is important to notice that the third term $-\gamma\ln(r_v/r_r)$ in the right-hand side of (\ref{def_ns_Lambda_v}) becomes infinite when $r_v$ tends to $0$, leading to an ill-defined dry-air version of $\Lambda_v$.
However, the contribution to the moist squared BVF is proportional to the product of $\Lambda_v$ by the gradient $\partial q_v/\partial z$.
Accordingly, the dry-air limit must be computed within a given dry region around $q_0=q_v(z_0)=0$ and where $q_v(z)-q_0$ is positive and very close to zero, leading to a first order formulation of $q_v(z)$ proportional to $(z-z_0)^2$, and to a gradient proportional to $z-z_0$, and thus to $\sqrt{q_v(z)}$.
This result shows that the product $\Lambda_v \: (\partial q_v/\partial z)$ contains a term varying as $\ln(r_v)\:\sqrt{q_v}$ which has $0$ as a limit when $q_v$ and $r_v=q_v/(1-q_v)$ tend to zero.

The computations needed for deriving (\ref{def_ns_dT_dqt_1}) and (\ref{def_ns_Lambda_v}) are rather long.
They have been obtained with the help of the properties $q_t=q_v$, $r_t=r_v$, $r_t=q_t/(1-q_t)$, $\partial r_t / \partial q_t = (r_t/q_t)^2$, $r_v=q_v(1+r_v)$, $\kappa\:\eta=\gamma$, $\delta=\eta-1$ and in particular with the identity
\begin{equation}
  \frac{q_v}{r_v}
  \; - \; \frac{1 + \delta \: q_v}{1 + \eta \: r_v}
  \; = \; 0
  \: . \label{def_prop_ns}
\end{equation}

From (\ref{def_ns_drho_dqt}) and (\ref{def_ns_dT_dqt_1}), the second partial derivative involved in the formulation  (\ref{def_new_N2_moist_6}) for the non-saturated version of $N_m^2$ is equal to
\begin{align}
  \left.
  \frac{\partial \rho}{\partial q_t}
  \right|_{p,s}
  & = \: - \:  \rho \:
\left[
  \frac{R_v}{R}
  -
  \frac{c_{pd}}{c_p}
  \left(
  \Lambda_r + \Lambda_v
  \right) \:
  \right]
  \: + \:
  \rho \: \left(\frac{T}{T_v}\right)
\:
  . 
\label{def_ns_drho_dqt_2}
\end{align}
In order to be consistent with the saturated version derived in the next Appendix,  (\ref{def_ns_drho_dqt_2}) can be written differently.
The last term $(T/T_v)$ is replaced by $(1+\eta\:r_v)\:(T/T_v)$ and the additional part $\eta\:r_v\:(T/T_v)$ is then subtracted from the bracketed term of (\ref{def_ns_drho_dqt_2}), together with the following identities 
\begin{align}
  \frac{R_v}{R} \: + \: \eta \: r_v \: \frac{T}{T_v}
 & = \;
 ( 1  +  r_v ) \:
 \frac{R_v}{R} 
\:  , \label{def_ns_prop3}
\\
  ( 1  +  \eta \: r_v ) \:
 \frac{T}{T_v} 
 & = \;
 ( 1  +  r_v ) 
 \; = \;
 \frac{1}{1-q_v} 
\:  . \label{def_ns_prop4}
\end{align}
   The result is
\begin{align}
  \left.
  \frac{\partial \rho}{\partial q_t}
  \right|_{p,s}
  & = \: - \:  \rho \:
\left[
   ( 1  +  r_v ) \: \frac{R_v}{R}
  -
  \frac{c_{pd}}{c_p}
  \left(
  \Lambda_r + \Lambda_v
  \right) \:
  \right]
 \: + \:
 \frac{\rho}{1-q_v} 
\:  .  \label{def_ns_drho_dqt_3}
\end{align}

%----------------- -------------------------------------
%    APPENDIX - D   
%----------------- -------------------------------------
%----------------------------------------------------------------------
\vspace{4mm}
\noindent
{\bf Appendix D. The saturated moist squared BVF.}
%----------------------------------------------------------------------
\label{appendix_sat_BVF}
\renewcommand{\theequation}{D.\arabic{equation}}
  \renewcommand{\thefigure}{D.\arabic{figure}}
  \renewcommand{\thetable}{D.\arabic{table}}
      \setcounter{equation}{0}
        \setcounter{figure}{0}
        \setcounter{table}{0}
\vspace{1mm}

The saturated squared BVF is  computed in this Appendix only using liquid water content, since the hypotheses retained in the Appendix~B do not allow the possibility of having liquid and solid species in a parcel of moist air at the same time .
In fact, the same hypothesis is made for the derivation of the specific moist entropy formulation (\ref{def_S_THs}), with the sum $L_{vap}\:q_l + L_{sub}\:q_i $ to be understood as $L_{vap}\:q_l$ or $L_{sub}\:q_i $, depending on $T>0$ or $T<0$, with either $q_l\neq0$ or $q_i\neq0$, respectively.
The ice content formula can be derived through the symmetry properties: $L_{vap}$ replaced by $L_{sub}$, $q_l$  by $q_i$ and $r_l$ by $r_i$.

The properties $q_i=0$,  $q_v=q_{sw}(T,p,q_t)$, and $q_l=q_t-q_{sw}(T,p,q_t)$ are used to derive the following saturated moist air version of the state equation (\ref{def_eq_state}), to arrive at
\begin{align}
\rho (T, q_t, p)  &  \: = \;
    \frac{1}{T} \;
    \frac{1}{1+\eta\:q_{sw}(T,p,q_t) - q_t} \;
    \frac{p}{R_d}
    \: , \label{def_rho_sat}
\\
q_{sw} (T,q_t,p) & = \:
      \frac{\varepsilon \; e_{sw}(T) }{p-e_{sw}(T)}\:(1-q_t)
     \: , \label{def_prop1}
\\
\rho (T, q_t, p)  &  \: = \;
    \frac{p-e_{sw}(T)}{R_d\:T\:(1-q_t)}
    \: . \label{def_rho_sat2}
\end{align}
The liquid water saturated entropy can be written as
\begin{align}
  s (T,q_t,p)
& \: = \; s_{ref}
 \: + {c}_{pd} \: \ln\left( T \right)
 \: + {c}_{pd} \: \lambda \: q_t \: \ln\left( \frac{T}{T_r} \right)
 \: - R_d \: \ln\left( \frac{p}{p_0} \right)
\nonumber \\
& \quad
    + \: \frac{L_{vap}}{T} \: q_{sw}
  \: - \: {c}_{pd} \: \left( \frac{L_{vap}}{{c}_{pd}\:T}
                            \: - \: \Lambda_r  \right)  \: q_t
  \: - {c}_{pd} \: \kappa \: \delta \: q_t \: \ln\left( \frac{p}{p_r} \right)
\nonumber \\
& \quad
    - {c}_{pd} \: \gamma \: q_t \: \ln\left( \frac{r_{sw}}{r_r} \right)
 \: + {c}_{pd} \: \kappa \: \delta \: q_t \:
                \ln\left( \frac{1+\eta\:r_{sw}}
                                {1+\eta\:r_r}
                    \right)
 \: + {c}_{pd} \: \kappa \:
                \ln\left( 1+\eta\:r_{sw} \right)
    \, ,
  \label{def_S_sat}
\end{align}
where $L_{vap}$ only depends on $T$ and where, from (\ref{def_prop1}),  $r_{sw}=q_{sw}/(1-q_t)$ only depends on $T$ and $p$.

The computations of the partial derivatives of $\rho$ with respect to $s$ or $q_t$ are more complicated than for the unsaturated cases and the following properties must be taken into account:
\begin{align}
\frac{1}{e_{sw}} \frac{d e_{sw}}{dT} & = \:
      \frac{L_{vap}(T) }{R_{v}\:T^2}
      \: , \label{def_des_dT}
\\
c_{pd} \: ( 1 + \lambda \: q_t )
+ (c_l-c_{pv}) \: q_l
  & \: = \: 
  c_p
  \: , \label{def_prop_sat}
\\
     L_{vap}(T) + \left( c_l-c_{pv} \right) T  & \: = \: L_{v}^0 \: = \: C^{ste}
\label{def_Lv_0} \: ,
\\
 - \:  T^2 \: \frac{\partial }{\partial T}
       \left( \frac{L_{vap}(T)}{T} \right)
& =  \: L_{v}^0 
      \: , \label{def_dLv_dT}
\\
\frac{p}{p-e_{sw}(T)} 
& = \: 1+\eta\:r_{sw}(T,p)
    \; , \label{def_esw}
\end{align}
\begin{align}
  \frac{q_t}{r_{sw}}
  \: - \: 
  \frac{1 + \delta \: q_t}{1 + \eta \: r_{sw}}
  & = \;
  \frac{q_l}{r_{sw}\:(1 + \eta \; r_{sw})}
  \: , \label{def_prop5}
\\
   ( 1 + \eta \: r_{sw} ) \: \frac{T}{T_v}
& = \:
   \frac{1}{1-q_t}
 \: , \label{def_prop4}
\\
 \frac{p-e_{sw}}{p\:(1-q_t)}
   \; = \:
  \frac{T}{T_v}
&   = \:
  \frac{R_d}{R}
 \; = \:
\frac{1}{1 + \eta \; q_{sw} - q_t}
  \:  . \label{def_prop3}
\end{align}
The derivative at constant pressure of the saturating specific content and mixing ratio are equal to
\begin{align}
  \left.
  \frac{\partial q_{sw}(T,q_t,p)}{\partial T}
  \right|_{p,q_t}
& = \: \left( 1 + \eta \: r_{sw} \right) \:
      \frac{L_{vap} \: q_{sw}}{R_{v}\:T^2}
  \: , \label{def_dqs_dT}
\\
  \left.
  \frac{\partial r_{sw}(T,p)}{\partial T}
  \right|_{p,q_t}
& = \: \left( 1 + \eta \: r_{sw} \right) \:
      \frac{L_{vap} \: r_{sw}}{R_{v}\:T^2}
  \: . \label{def_drs_dT}
\end{align}
Moreover, chain rules are applied to the derivatives of $q_{sw}(T,q_t,p)$ and $r_{sw}(T,p)$, leading to
\begin{align}
  \left.
  \frac{\partial q_{sw}(T,q_t,p)}{\partial s}
  \right|_{p,q_t}
  & =  \:
  \left.
  \frac{\partial q_{sw}}{\partial T}
  \right|_{p,q_t}
  \left.
  \frac{\partial T}{\partial s}
  \right|_{p,q_t}
  , \label{def_sat_chain_1} \\
  \left.
  \frac{\partial r_{sw}(T,p)}{\partial s}
  \right|_{p,q_t}
  & =  \:
  \left.
  \frac{\partial r_{sw}}{\partial T}
  \right|_{p}
 \:
  \left.
  \frac{\partial T}{\partial s}
  \right|_{p,q_t}\!
  . \label{def_sat_chain_2}
\end{align}

The previous properties allow the computation of the first partial derivative of (\ref{def_rho_sat}) with respect to $s$ (at constant values for $p$ and $q_t$), leading to
\begin{align}
  \left.
  \frac{\partial \rho}{\partial s}
  \right|_{p,q_t}
  & = \: - \: \rho \; D_{1w} \;
  \frac{1}{T} \;
  \left.
  \frac{\partial T}{\partial s}
  \right|_{p,q_t}
  \: , \label{def_sat_drho_ds}
\end{align}
where $D_{1w}$ is given by (\ref{def_sat_D1l}).

The partial derivative of $T$ with respect to $s$ (at constant values for $p$ and $q_t$)  is  obtained by computing the derivative of (\ref{def_S_sat}) with respect to $s$ and with $T=T(s,q_t,p)$, to arrive at
\begin{align}
  \left.
  \frac{\partial s}{\partial s}
  \right|_{p,q_t}
  & \: \equiv \; 1 \: = \; D_{2w} \;
  \frac{c_p}{T} \;
  \left.
  \frac{\partial T}{\partial s}
  \right|_{p,q_t}
  \: , \label{def_sat_dT_ds_1}
\end{align}
where $ D_{2w}$ is given by  (\ref{def_sat_D2l}).

From (\ref{def_sat_drho_ds}) and (\ref{def_sat_dT_ds_1}), the first partial derivative  involved in the liquid-water saturated formulation  (\ref{def_new_N2_moist_6}) for $N_m^2$ is equal to
\begin{equation}
  \left.
  \frac{\partial \rho}{\partial s}
  \right|_{p,q_t}
  \: = \: - \: \frac{\rho}{c_p}
              \: \frac{D_{1w}}{D_{2w}}
  \: . 
\label{def_sat_drho_ds_2}
\end{equation}
The difference with the non-saturated case (\ref{def_ns_drho_ds_2}) is the extra term $({D_{1w}}/{D_{2w}})$.

The second partial derivative of $\rho$ with respect to $q_t$ (at constant values for $p$ and $s$) is computed from (\ref{def_rho_sat}) with $T=T(s,q_t,p)$ and with the chain rule applied to the derivative of $q_{sw}(T,q_t,p)$,  leading to
\begin{align}
  \frac{1}{\rho}
  \left.
  \frac{\partial \rho}{\partial q_t}
  \right|_{p,s}
   & = \: - \:
  \frac{D_{1w}}{T} \;
  \left.
  \frac{\partial T}{\partial q_t}
  \right|_{p,s}
 \: + \:
   \frac{1}{1-q_t}
  \: . \label{def_sat_drho_dqt}
\end{align}
The term $D_{1w}$ is  again given by (\ref{def_sat_D1l}).

The partial derivative of $T$ with respect to $q_t$ (at constant values for $p$ and $s$) is obtained by computing the derivative of (\ref{def_S_sat}) with respect to $q_t$, with $T=T(s,q_t,p)$ and with chain rules applied to $q_{sw} (T,q_t,p)$ and $r_{sw} (T,p)$, leading to
\begin{align}
  \left.
  \frac{\partial q_{sw}}{\partial q_t}
  \right|_{p,s}
& = \: ( 1 + \eta \: r_{sw} ) \:
      \frac{L_{vap} \: q_{sw}}{R_{v}\:T^2}
  \left.
  \frac{\partial T}{\partial q_t}
  \right|_{p,s} \!
 - \: r_{sw}
  \: , \label{def_dqs_dqt}
\\
  \left.
  \frac{\partial s}{\partial q_t}
  \right|_{p,s}
& \equiv \; 0 
  \: = \: D_{2w} \;
  \frac{c_p}{T}
  \left.
  \frac{\partial T}{\partial q_t}
  \right|_{p,s} \!
  - \:
  (1 +  r_{sw}) \:
  \frac{L_{vap}}{T}
  \; + \:
  c_{pd} \:
  \left(\:
  \Lambda_r + \Lambda_{sw}
  \:\right)
  \: . \label{def_sat_dT_dqt_1}
\end{align}
The term $D_{2w}$ is again given by (\ref{def_sat_D2l}).
The term  $\Lambda_{sw}$ is equal to the non-saturated version of  $\Lambda_v$ given by (\ref{def_ns_Lambda_v}), but expressed for the saturated conditions $r_v=r_{sw}$, leading to (\ref{def_sat_Lambda_s}).

Even if the results are compact and consistent at first sight with (\ref{def_ns_dT_dqt_1}) and (\ref{def_ns_Lambda_v}), the computations made to derive (\ref{def_sat_dT_dqt_1})  and (\ref{def_sat_Lambda_s}) are rather long.
They have been obtained by taking into account the properties $q_i=0$, $q_v=q_{sw}(T,q_t,p)$, $q_t=q_{sw}+q_l$, together with the results (\ref{def_prop1}) to (\ref{def_drs_dT}).

From (\ref{def_sat_drho_dqt}) and (\ref{def_sat_dT_dqt_1}), the second partial derivative  involved in the formulation  (\ref{def_new_N2_moist_6}) for $N_m^2$ is thus equal to
\begin{align}
  \left.
  \frac{\partial \rho}{\partial q_t}
  \right|_{p,s} \!\!
  & =
  - \: \rho
    \: \frac{D_{1w}}{D_{2w}}
  \left[ \,
    (1+r_{sw})\:
    \frac{L_{vap}}{c_p\:T}
  - 
    \frac{c_{pd}}{c_p}
   \left(
      \Lambda_r + \Lambda_{sw}
    \right) \,
    \right]
  \; +
  \frac{\rho}{1-q_t}
  \: .  \label{def_sat_drho_dqt_2}
\end{align}

%----------------- -------------------------------------
%    APPENDIX - E  
%----------------- -------------------------------------
%----------------------------------------------------------------------
\vspace{4mm}
\noindent
{\bf Appendix E. Comparisons with the buoyancy formulations.}
%----------------------------------------------------------------------
\label{appendix_N2_Theta_v}
\renewcommand{\theequation}{E.\arabic{equation}}
  \renewcommand{\thefigure}{E.\arabic{figure}}
  \renewcommand{\thetable}{E.\arabic{table}}
      \setcounter{equation}{0}
        \setcounter{figure}{0}
        \setcounter{table}{0}
\vspace{1mm}

The specific moist entropy formulations (\ref{def_BVF_ns_new}) and  (\ref{def_BVF_sat_new_l})   have been obtained without any approximation, except the hydrostatic one used in the Appendix~B to demonstrate the generic BFV formula (\ref{def_new_N2_moist_6}).

The advantage of the formulations (\ref{def_BVF_ns_new})-(\ref{def_BVF_sat_new_l}) is that they are expressed in terms of the gradients of the more general conservative variables, i.e. the specific moist entropy and the chemical fractions of the parcel (or equivalently the concentrations in dry air or total water content).

It is  generally accepted that the non-saturating squared BVF  may be defined by
\begin{align}
  N^2_{v}
  & = \frac{g}{\theta_v}
    \: \frac{\partial \theta_v }{\partial z}
  \: .
\label{def_BVF_N2v_ns}
\end{align}
It may be important to try to express the specific moist entropy formulations  (\ref{def_BVF_ns_new})  and (\ref{def_BVF_sat_new_l})  in terms of the vertical gradient of the buoyancy potential temperature  $\theta_v$ given by  (\ref{def_Tv2}).

Let us derive the non-saturated moist air formula (\ref{def_BVF_ns_new}) in terms of the vertical gradient of $q_v=q_t$ and  $\theta_v$ by computing both the vertical gradient of $\theta_v$ and the vertical gradient of the specific moist entropy  (\ref{def_S_THs}), with $\theta_s$ given by (\ref{def_THs}) and with the hydrostatic assumption, leading to
\begin{align}
\frac{g}{\theta_v}
    \: \frac{\partial \theta_v }{\partial z}
  & \; = \; \frac{g}{\theta}
    \: \frac{\partial \theta }{\partial z}
\: + \: g\:
    \delta \:
    \: \frac{T }{T_v}
    \: \frac{\partial q_v }{\partial z}
  \: ,
\label{def_grad_theta_v} \\
g\: \frac{c_{pd}}{c_p}
    \: \frac{\partial \ln(\theta_s) }{\partial z}
  & \; = \; \frac{g}{\theta}
    \: \frac{\partial \theta }{\partial z}
\: + \: \frac{g \:c_{pd}}{c_p}
    \left(
    \Lambda_r + \Lambda_{v}
  \right)
    \: \frac{\partial q_v }{\partial z}
\; - \: \frac{g^2 \: q_v}{c_p \: T_v }
    \left(
    \lambda - \delta
  \right)
  \: .
\label{def_grad_theta_s}
\end{align}

When (\ref{def_grad_theta_s})  and (\ref{def_grad_theta_v})  are inserted into the non-saturated moist air formula (\ref{def_BVF_ns_new}), most of the terms cancel out and the final result is
\begin{align}
  N^2_{ns}
  & \; = \: \frac{g}{\theta_v}
\left(
    \: \frac{\partial \theta_v }{\partial z}
  \: - \: \Gamma_{c/ns} \:
\right)
  .
\label{def_BVF_N2v_ns_CG}
\end{align}
The comparison of (\ref{def_BVF_N2v_ns}) with (\ref{def_BVF_N2v_ns_CG}) shows that a non-saturated counter-gradient term $\Gamma_{c/ns}$ appears.
It is equal to
\begin{align}
\Gamma_{c/ns}
  & \; = \:
    \frac{g}{c_p} \;
    \frac{\theta}{T} \:
    \left(
    \lambda - \delta
  \right)
\: q_v 
  \: .
\label{def_BVF_N2v_ns_CG2}
\end{align}
It depends on $q_v$ and, for a typical moist PBL where $q_v=10$~g~kg${}^{-1}$, $\theta \approx T$ and $c_p \approx 1000$~J~K${}^{-1}$~kg${}^{-1}$, the values $\lambda - \delta \approx 0.23$ leads to $\Gamma_{c/ns} \approx 0.023$~K~km${}^{-1}$.
This implies a contribution of $-0.008$~$10^{-4}$~s${}^{-2}$ to $N^2_{ns}$.
It is a small term when it is compared to significant values of $N^2_{ns}$ which are typically about $100$ times larger.

%----------------- -------------------------------------
%    APPENDIX - F  
%----------------- -------------------------------------
%----------------------------------------------------------------------
\vspace{4mm}
\noindent
{\bf Appendix F. Transitions betwen unsaturated and saturated formulations.}
%----------------------------------------------------------------------
\label{appendix_FC_MC}
\renewcommand{\theequation}{F.\arabic{equation}}
  \renewcommand{\thefigure}{F.\arabic{figure}}
   \renewcommand{\thetable}{F.\arabic{table}}
      \setcounter{equation}{0}
        \setcounter{figure}{0}
         \setcounter{table}{0}
\vspace{1mm}

When comparing both unsaturated and saturated moist expressions for the squared BVF, one notices a good deal of similarity: the ``water lifting term'' is the same, the multiplying gradients are identical and there are thus only three kinds of transitions to consider. The first one is the natural expression of the terms depending on $r_v$ in the unsaturated case by the equivalent in terms of $r_{sw}$ in the saturated case, this being applied to both the $(1+r_v)$ multiplicator and to the $\Lambda_v$ expression. The second one is the transition between lapse rates $\Gamma_{ns}$ and $\Gamma_{sw}$, i.e. the multiplication by $D_{1w}/D_{2w}$. The third one is the replacement of $(c_p\:R_v)/R$ by $L_{vap}/T$.

The crucial point for ensuring a complete and smooth transition between the two formulations is that, together with the logical number one transition, going from $D_{1w}/D_{2w}$ to $1$ and from $L_{vap}/T$ to $(c_p\:R_v)/R$ {\em both\/} just require replacing $(L_{vap}\:R)/(c_p\:R_v\:T)$ by $1$!

A simple way to create a generalized $N^2$ formula on the basis of this structural symmetry between both transitions is therefore to define
\begin{align}
F(C) & = \:
1 + C\: 
\left[\:
  \frac{L_{vap}}{c_p\:T}
  \frac{R}{R_v}
  \: - \: 1 \:
\right]_E \: ,
\label{def_FC} \\
D_C & = \:
\left(
  \frac{L_{vap}\:r_{sw}}{R_d\:T}
\right)_E \: ,
\label{def_DC} \\
 M(C) & = \:
\frac{
1 + D_C \:
}
{
  1 
  +
  D_C \:
  F(C) 
}
 \: ,
\label{def_MC}
\end{align}
for obtaining
\begin{align}
  &N^2(C)
  \: = \: g \: \frac{c_{pd}}{c_p} \: M(C)
     \left(
      \frac{\partial \ln(\theta_s) }{\partial z}
     \right)_E
+ \: g\: 
     \left(
      \frac{\partial \ln(q_d)  }{\partial z}
     \right)_E
\nonumber \\
  & \quad \quad + \: g \; M(C) \; F(C) \:
   \left[ \:
    (1+r_v)\: \frac{R_v}{R} \:
   \right]_E
     \left(
      \frac{\partial q_t  }{\partial z}
     \right)_E
\nonumber \\
  & \quad \quad - \: g \: \frac{c_{pd}}{c_p} \:  M(C) \:
   \left[
     \: \Lambda_r+\Lambda_v \:
   \right]_E
     \left(
      \frac{\partial q_t  }{\partial z}
     \right)_E
  .
\label{def_BVF_FC_MC}
\end{align}

In the above set of equations the subscript ``E'' represents the environmental value of any moist air parcel, independently whether the conditions are fully unsaturated, partly saturated or fully saturated.

Even if the exact identity between both analytical manifestations of the transition via $F(C)$ is a welcome result, there is some physical consistency in having the same term playing the key role in the lessening of the resistance to buoyant motions when condensation occurs on the one hand and in the replacement within the buoyancy term of gaseous density effects by latent heat release impacts on the other hand.

\begin{figure}[hbt]
\centering
 \includegraphics[width=0.49\linewidth,angle=0,clip=true]{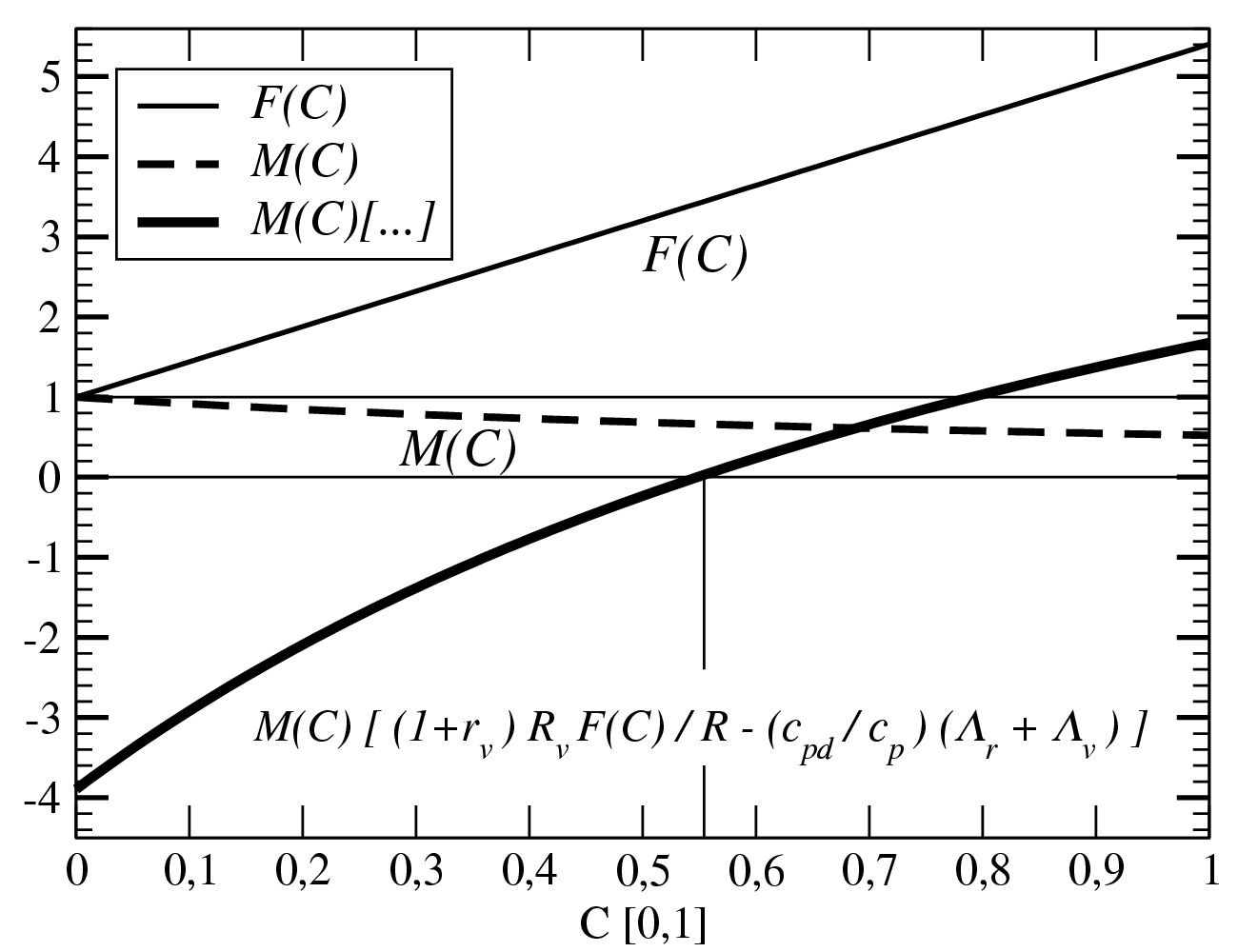}
\caption{\it\small
The curves $F(C)$ and $M(C)$, defined by (\ref{def_FC}) to (\ref{def_MC}), are computed for the ``just-saturated'' conditions ($T=10$~C, $p=900$~hPa, $q_v=q_{sw}$, $q_l=0$) and for $C$ varying from $0$ (unsaturated) to $1$ (saturated).
The values $0$ and $1$ are depicted as thin horizontal and solid lines.
The third curve $M(C)\:[\:(1+r_v)\:R_v\: F(C)/R - \: ({c_{pd}/c_p}) \: (\Lambda_r+\Lambda_v)\:]$ corresponds to the new extra term in the second line of (\ref{def_BVF_FC_MC}), with a thin line added to show that the zero value is obtained for $C_0 \approx 0.55$.
\label{fig_FC_MC}}
\end{figure}

It can be verified that the two formulas (\ref{def_BVF_ns_new}) and (\ref{def_BVF_sat_new_l}) are obtained from (\ref{def_BVF_FC_MC}) by the two limit cases described in the Table~\ref{Table_FC_MC}. 
Furthermore the product of $F(C)$ by $M(C)$ within the last but one  term clearly shows that we have here more than a linear interpolation of $N^2$ between the two extreme cases, as indicated by the non-linear heavy solid curve depicted in Figure~\ref{fig_FC_MC}.

In fact, the special regime that cancels the second and third  lines of (\ref{def_BVF_FC_MC}) corresponds to the property
 \begin{align}
 C_0
 \left(
     \frac{L_{vap}}{c_p\:T}
     \:
     \frac{R}{R_v} 
     - 1
 \right)
  &  = \: \frac{c_{pd}}{c_p} 
     \left(
     \frac{\Lambda_r+\Lambda_v}{1+r_v}
     \right)
     \frac{R}{R_v} 
     - 1
\label{def_C0} 
\: .
\end{align}

For the typical values $\Lambda_r \approx 5.87$, $\Lambda_v \approx -0.32$, $R/R_v \approx 0.622$ and  $T=283$~K, then $L_{vap}/(c_p\:T) \approx 8.6$ and $(\Lambda_r+\Lambda_v)/(1+r_v) \approx 5.5$, leading to the reversal value $C_0\approx (5.5\times0.622-1)/(8.6\times0.622-1) \approx 0.55$.
This value of $C_0$ can be associated with the approximate $5.5/8.6 \approx 2/3$ position of ${\theta}_{s}$ between $\theta_l$ and $\theta_E$, as seen in Figures~\ref{fig_THS1_part1} and \ref{Fig_RF03B_N2_results}(a).

%===============
% Figure (F.1) : sous FIRE_I / zprog_calcul_9b_N2MG_review / ITRACE_N2_FC_MC ; ICONVERT3
%===============

Lastly, on top of the contribution for obtaining $\theta_s$, (\ref{def_BVF_FC_MC}) clearly separates the roles of $q_t$ and $q_v$ in the rest of the $N^2$ expression. 
The total content $q_t$ (or its complement to one $q_d$) is present only through its vertical gradient, a key quantity in our way of obtaining the squared BVF.
The water-vapour content $q_v$ is present only in the second- and third-lines parenthesis via the moist definition of $c_p$, $(1+r_v)$ and $\Lambda_v$, under the implicit understanding that it will equal $q_{sw}$ for $C=1$ in all these occurrences.
One should nevertheless realize that, numerically speaking, the above implicit assumption may not be perfectly obeyed without this having any bad consequences for the computation of a generalized moist squared BVF according to Equation (\ref{def_BVF_FC_MC}).

%%%%%%%%%%%%%%%
%% TABLE F.1 %%
%%%%%%%%%%%%%%%
% \toprule  \midrule \bottomrule
\begin{table}
\caption{\it\small
The values for $F(C)$ and $M(C)$ for the unsaturated case defined by $C=0$ and ($q_v$, $r_v$, $\Lambda_v$), and for the saturated case defined by $C=1$ and ($q_{sw}$, $r_{sw}$, $\Lambda_{sw}$). The moist formulations of $c_p$ and $R$ slightly depend on $q_d=1-q_t$, $q_v$, $q_l$ or $q_i$.
\label{Table_FC_MC}}
\vspace*{2mm}
\centering
\begin{tabular}{|c||c|c|}
\hline
         & $C=0$ & $C=1$ \\ 
\hline   \hline  
    $F(C)$  & $1$ & $(L_{vap}\:R)/(c_p\:R_v\:T)$  \\ 
\hline
    $M(C)$  & $1$ &  $D_{1w}/D_{2w}$   \\
\hline
\end{tabular}
\end{table}

There are however some caveats associated with the use of $N^2(C)$.
First, it is clear that except in the extreme homogeneous cases corresponding to $C=0$ and to $C=1$, the squared BVF  looses its original meaning associated with the period of natural oscillations of an air parcel displaced along the vertical.
Second, it is no longer possible to find an equivalent of (\ref{def_new_N2_moist_6}) leading to (\ref{def_BVF_FC_MC}), since the simplifications allowing us to express $T$ as a function of $s$, $q_t$ and $p$ (as used in Appendices~C and D) have no equivalent in the case of a grid-mesh partly saturated and partly unsaturated.

Nevertheless, $N^2(C)$ has the physical dimension of a squared BVF, and its expression closely follows, term by term, the physical logic explained in the sections~6 and 7.
It is thus our belief that it might be used in some applications, provided that the meaning of $C$ is not over-interpreted.

Furthermore, by construction, the shape of this function is linked to the important issue already discussed of the change of sign of the terms in the second lines of (\ref{def_BVF_ns_new}) and (\ref{def_BVF_sat_new_l}). 
It is expected that the unstable (stable) feature of isentropic motions of moist air, which is in correspondence with saturated (unsaturated) condition, must also have a neutral case in between.
Even if the definition of a transition parameter remains a complex issue, well beyond the scope of the present work, $N^2(C)$ offers a monotonic path leading to some ``moist neutrality'' for a value of $C$ within the interval $[0,1]$.

%\end{document}

%-------------------------------------------------------------------------
%    REFERENCES
%-------------------------------------------------------------------------

%\begin{thebibliography}

%\newpage

\vspace{5mm}
\noindent{\large\bf References}
\vspace{2mm}

\noindent{$\bullet$ Betts AK.} {1973 (B73)}.
{Non-precipitating cumulus convection and its parameterization.
{\it Q. J. R. Meteorol. Soc.}
{\bf 99} (419):
178--196.}

\noindent{$\bullet$ Bolton D.} {1980}.
{The computation of Equivalent Potential Temperature.
{\it Mon. Weather Rev.}
{\bf 108,} (7):
1046--1053.}

\noindent{$\bullet$ Durran DR, Klemp JB.} {1982 (DK82)}.
{On the effects of moisture on the Brunt-V\"{a}is\"{a}l\"{a} Frequency.
{\it J. Atmos. Sci.}
{\bf 39} (10):
2152--2158.}

\noindent{$\bullet$ Emanuel KA.} {1994 (E94)}.
{Atmospheric convection.}
Pp.1--580.
Oxford University Press: New York and Oxford.

\noindent{$\bullet$ Geleyn J-F, Marquet P.} {2010}.
{Moist thermodynamics and moist turbulence for modelling at the non-hydrostatic scales.}
Pp.55-66.
ECMWF Workshop on Non-hydrostatic modelling, Reading, 8th of November, 2010.

\noindent{$\bullet$ Lalas DP, Einaudi F.} {1974 (LE74)}.
{On the correct use of the wet adiabatic lapse rate in stability 
criteria of a saturated atmosphere.
{\it J. Appl. Meteor.}
{\bf 13} (3):
318--324.}

\noindent{$\bullet$ Lilly DK.} {1968}.
{Models of cloud-topped mixed layers under a strong inversion.
{\it Q. J. R. Meteorol. Soc.}
{\bf 94} (401):
292--309.}

\noindent{$\bullet$ Marquet P.} {2011 (M11)}.
{Definition of a moist entropic potential temperature. Application to FIRE-I data flights.
{\it Q. J. R. Meteorol. Soc.}
{\bf 137} (656):
768--791.
\url{http://arxiv.org/abs/1401.1097}.
{\tt arXiv:1401.1097 [ao-ph]}}

\noindent{$\bullet$ Pauluis O., Held I.M.} {2002}.
{Entropy budget of an atmosphere in radiative-convective equilibrium. Part I:
maximum work and frictional dissipation.
{\it J. Atmos. Sci.}
{\bf 59} (2):
125--139.}

\noindent{$\bullet$ Pauluis O.} {2008 (P08)}.
{Thermodynamic consistency of the anelastic approximation for a moist atmosphere.
{\it J. Atmos. Sci.}
{\bf 65} (8):
2719--2729.}

\noindent{$\bullet$ Pauluis O., Czaja A., Korty R.} {2010a (PCK10)}.
{The global atmospheric circulation in moist isentropic coordinates.
{\it J. Climate.}
{\bf 23} (11):
3077--3093.}

\noindent{$\bullet$ Pauluis O., Schumacher J.} {2010b (PS10)}.
{Idealized moist Rayleigh-B\'enard convection with piecewise
linear equation of state.
{\it Commun. math. Sci.}
{\bf 8} (1):
295--319.}

\noindent{$\bullet$ Pauluis O.} {2011 (P11)}.
{Water vapor and mechanical work: a comparison of Carnot and Steam cycles.
{\it J. Atmos. Sci.}
{\bf 68} (1):
91--102.}

\noindent{$\bullet$ Ruprecht D., Klein R., Majda A.J.} {2010}.
{Modulation of Internal Gravity Waves in a Multi-scale Model for Deep Convection on Mesoscales.
{\it J. Atmos. Sci.}
{\bf 67} (8):
2504--2519.}

\noindent{$\bullet$ Ruprecht D., Klein R.} {2011}.
{A Model for nonlinear interactions of internal gravity waves with saturated regions.
{\it Met. Zeitschr.}
{\bf 20} (2):
243--252.}

\end{document}